\documentclass[fleqn,usenatbib]{mnras}
\usepackage{newtxtext,newtxmath}
\usepackage[T1]{fontenc}
\DeclareRobustCommand{\VAN}[3]{#2}
\let\VANthebibliography\thebibliography
\def\thebibliography{\DeclareRobustCommand{\VAN}[3]{##3}\VANthebibliography}

\usepackage{graphicx}	% Including figure files
\usepackage{amsmath}
\usepackage{amssymb}	% Extra maths symbols
\usepackage{multicol}        % Multi-column entries in tables
\usepackage{bm}		% Bold maths symbols, including upright Greek
\usepackage{pdflscape}	% Landscape pages
\usepackage{threeparttable}
\usepackage{multirow}
\defcitealias{Liu2020}{Paper I}
\defcitealias{Wang2022}{Paper II}
\newcommand{\kms}{km\,$\rm s^{-1}$}

\title[Probing the Galactic halo with RR Lyrae stars ]{Probing the Galactic halo with RR Lyrae stars $-$ III. The chemical and kinematic properties of the stellar halo}

\author[Liu et al.]{
Gaochao Liu,$^{1,2,3}$\thanks{E-mail: gcliu@ctgu.edu.cn}
Yang Huang,$^{4,5}$\thanks{E-mail: huangyang@bao.ac.cn}
Sarah Ann Bird,$^{1,2}$
Huawei Zhang,$^{6,7}$ 
Fei Wang$^{6,7}       $
 and Haijun Tian$^{1,2,8}$
\\
$^{1}$College of Science, China Three Gorges University, Yichang 443002, People's Republic of China\\
$^{2}$Center for Astronomy and Space Sciences, China Three Gorges University, Yichang 443002, People's Republic of China\\
$^{3}$ Key Laboratory of Quark and Lepton Physics (MOE), Central China Normal University, Wuhan 430079, People's Republic of China\\  
$^{4}$ University of Chinese Academy of Sciences, Beijing 100049,  People's Republic of China\\
$^{5}$ National Astronomical Observatories, Chinese Academy of Sciences, Beijing 100012, People’s Republic of China\\
$^{6}$Department of Astronomy, Peking University, Beijing 100871, People's Republic of China\\
$^{7}$Kavli Institute for Astronomy and Astrophysics, Peking University, Beijing 100871, People's Republic of China\\
$^{8}$School of Physics and Astronomy, China West Normal University, NanChong 637002, People’s Republic of China
}

\date{Accepted XXX. Received YYY; in original form ZZZ}

\pubyear{2022}

\begin{document}
\label{firstpage}
\pagerange{\pageref{firstpage}--\pageref{lastpage}}
\maketitle

\begin{abstract}
 Based on a large spectroscopic sample of $\sim$ 4,300 RR Lyrae stars with metallicity, systemic radial velocity and distance measurements, we present a detailed analysis of the chemical and kinematic properties of the Galactic halo. Using this sample, the metallicity distribution function (MDF) as a function of $r$ and the velocity anisotropy parameter $\beta$ profiles (for different metallicity populations) are derived for the stellar halo. Both the chemical and kinematic results suggest that the Galactic halo is composed of two distinct parts, the inner halo and outer halo. 
The cutoff radius ($\sim$ 30 kpc) is similar to the previous break radius found in the density distribution of the stellar halo. We find that the inner part 
is dominated by a metal-rich population with extremely radial anisotropy ($\beta \sim 0.9$). These features are in accordance with those of ``{\it Gaia}-Enceladus-Sausage'' (GES) and we attribute this inner halo component as being dominantly composed of stars deposited from this ancient merged satellite. We find that GES probably has a slightly negative metallicity gradient. The metal-poor populations in the inner halo are characterized as a long-tail in MDF with an anisotropy of $\beta \sim 0.5$, which is similar to that of the outer part. The MDF for the outer halo is very broad with several weak peaks and the value of $\beta$ is around 0.5 for all metallicities.
\end{abstract}


%%%%%%%%%%%%%%%%%%%%%%%%%%%%%%%%%%%%%%%%%%%%%%%%%%

%%%%%%%%%%%%%%%%% BODY OF PAPER %%%%%%%%%%%%%%%%%%

\section{Introduction} \label{sec:intro}

Understanding the formation and evolution of galaxies is one of the most challenging projects in modern astrophysics.
The vast quantity of Galactic stars which we can study with 7D information (3D velocities, 3D positions and metallicity) is incomparable with any other galaxies.
As one of the oldest components of the Galaxy, the stellar halo preserves
information about the Galaxy's formation, evolution, and complex structures \citep[e.g.,][]{Helmi1999MNRAS, Chiba2000, Re_Fiorentin2005, Re_Fiorentin2015, Helmi2006, Helmi2017, Helmi2018, Kepley2007, Klement2008, Klement2009, Klement2011,  Morrison2009, Smith2009, Yuan_zhen2018, Koppelman2019},
and thus it is of vital importance to study this component.

\citet{Eggen1962} suggested that the oldest stars in the Galactic halo were formed during an ancient gravitation collapse. However, later observations show that the metallicities of Galactic globular clusters are independent of Galactocentric distances $r$, and
\citet{Searle1978} therefore proposed that the halo is formed through the merging of dwarf galaxies. 
Modern data from large-scale surveys show that the stellar halo has a complex structure
with multiple components and unrelaxed substructures, and continues to accrete matter from 
smaller dwarf galaxies which are then tidally disrupted in the Galaxy's gravitational field \citep[e.g.][]{Ibata1997, Belokurov2006, Belokurov2018, Schlaufman2009}. 
This confirms the prediction of the hierarchical galaxy formation model. Because of the long dynamical
timescales in the halo, tidal tails, shells, and other overdensities arising from accreted dwarf
galaxies remain observable over Gyrs, thus constituting a fossil record of the Milky Way's
accretion history \citep[e.g.][]{Helmi1999Nature, Helmi1999MNRAS, Helmi2018}. 

By analyzing a large sample with well-determined atmospheric and kinematic parameters selected from the Sloan Digital Sky Survey \citep[SDSS,][]{York2000}, \citet{Carollo2007} put forward a dual halo model for the Galaxy. The inner halo was dominated by the essentially radial merger of some massive metal-rich clumps 
then followed by a stage of adiabatic compression. The outer halo formed through dissipationless chaotic merging of smaller subsystems within a 
pre-existing dark-matter halo. The star formation within these massive clumps drive the mean metallicity to higher abundances for the inner halo, but the outer-halo population may be assembled from relatively more metal-poor stars. 
The MDF of the inner-halo peaks at  [Fe/H] = $-1.6$\,dex, with tails extending to higher and lower metallicities, but the outer-halo peaks around [Fe/H] = $-2.2$\,dex, a factor of 4 lower than that of the inner-halo population.

On the other hand, the velocity anisotropy parameter $\beta$, characterizing the shape of the velocity ellipsoid (spherical, radial or tangential),
 was only well measured at the solar neighborhood with a
typical value of $0.5-0.7$ \citep[e.g.][]{Smith2009, Bond2010, Evans2016, Posti2018} before the first data release of the ESA {\it Gaia} mission \citep{GaiaCollaborationBrown2016}. Few (indirect/direct) measurement attempts
in the outer halo were with large uncertainties \citep[e.g.][]{Kafle2012, Deason2012, Deason2013}. Now with {\it Gaia} and large stellar spectroscopic surveys (e.g. LAMOST and SDSS), the
direct measurements of the 3D velocity dispersions and $\beta$ over large distances is possible. 
\citet{Bird2019, Bird2020} analyzed the anisotropy profile of the halo of the Galaxy by using
blue horizontal branch stars (BHBs) and K giants, and found that the value of $\beta$ depends on metallicity and is nearly constant within 20 kpc for each metallicity bin, beyond which the anisotropy profile gently declines although remains radially dominated. 

Using a sample of main sequence stars, \citet{Belokurov2018} demonstrate that the Milky Way halo contains a large number of metal-rich
stars on extremely eccentric orbits and $\beta$ even can reach to 0.9. Using an established
mass-metallicity relationship \citep{Kirby2013} as well as numerical
simulations of Galaxy formation, \citet{Belokurov2018b} argue that the observed chemistry and the drastic radial anisotropy of the halo stars ( i.e. the ``{\it Gaia} Sausage") are telltale signs of a major accretion event by a satellite with 
$M_{\rm {vir}} > 10^{10}\, \mathrm{M}_\odot$ around the epoch of the Galactic disc formation, between 8 and 11 Gyr ago.
\citet{Iorio2018} and \citet{Wegg2019} also demonstrate 
the bulk of the Galactic stellar halo between 5 and 30 kpc characteristically have strongly radial orbits, 
yielding estimates of $ \beta \sim 0.9$, which is the result of the dramatic radialization of the massive progenitor's orbit, 
amplified by the action of the growing disc. Such an early merger event has been demonstrated by a plethora of studies \citep[e.g.][]{Vincenzo2019, Deason2019.490, Fattahi2019, Koppelman2020}. \citet{Koppelman2018} use the stars identified kinematically from {\it Gaia} DR2 to explore the phase-space structure and find a ``blob" connected with this merger event. \citet{Helmi2018} name the merger as ``{\it Gaia}-Enceladus", and independently discovered the remains of this large satellite from chemical analysis. For simplicity, throughout this work we call this component as the ``{\it Gaia}-Enceladus-Sausage" (GES).

In this work, we show the chemical and kinematic properties of the Galactic halo and 
explore the GES independently by our RR Lyrae (RRL) sample stars \citep{Liu2020}. 
As described in the introduction of \citet{Liu2020}, RRL are the most natural probe of the halo of the Galaxy since they are intrinsically bright (can be observed at great distances) and their absolute magnitudes only weakly depend on the metallicity which makes them suitable for measuring the MDF along Galactocentric distance (the number of observed RRL are almost unbiased at various distances). At the same time, with the measurement of systemic radial velocity combined with the proper motion provided by {\it Gaia} EDR3 \citep{GaiaCollaboration2021}, we can obtain their 3D velocities and analyze their kinematic properties. We use the metallicity density map, MDF, and anisotropy $\beta$ from our RRL sample to systematically map the Galactic halo.

 This paper is the third in a series (\citealt[][hereafter \citetalias{Liu2020}]{Liu2020}, \citealt{Wang2022}) based on RRL to explore the formation and evolution of the stellar halo of our Galaxy. 
 The data used in the current work is briefly described in Section\,2.  The MDF and kinematic properties of RRL are provided in Section\,3.  Section\,4 presents a discussion and comparison with recent results from the literature.
 Finally, we summarize in Section\,5.
 
\section{Data}\label{sect:data}

\citetalias{Liu2020} has constructed a catalog of RRL with metallicity estimates and systematic radial velocity measurements, derived from the SDSS \citep{Yanny2009} and LAMOST \citep{Deng2012, Zhao2012, Liu2014} spectroscopic data along with the photometric data taken from recent large-scale photometric surveys, such as QUEST \citep{Vivas2004, Mateu2012, Zinn2014}, SDSS Stripe\,82 \citep{Watkins2009, Sesar2010, Suveges2012}, Catalina \citep{Drake2013a, Drake2014},  LINEAR \citep{Sesar2013}, etc. In this study, we further compile additional photometric surveys, such as {\it Gaia} DR2 \citep{Clementini2019}, Pan-STARRS1 \citep{Chambers2016, Sesar2017}, and ASAS-SN \citep{Jayasinghe2018} into a unique 206,664 RRL photometric data set, then cross match with  the spectra from LAMOST DR6 and SDSS DR12. Finally, we obtain 8,172 RRL stars with both photometric and spectroscopic information. We process the data in the same manner as \citetalias{Liu2020} to measure spectroscopic metallicity and radial velocity. For objects with individual single-exposure spectral-noise-to-ratio (SNR) greater than 10 and not affected by shock waves (diagnosed by the equivalent width of the Ca~{\sc ii} K ), we measure the metallicities by a least $\chi^2$ fitting technique and take  the weighted mean metallicity of the individual spectra as the final adopted metallicity values. Then we fit the templates of systemic velocity of RRL provided by \citet{Sesar2012} to the observed spectra and derive the systemic radial velocity of our RRL sample. In total, metallicities for 7,324 RRL are obtained, 4,924 of them with systematic radial velocity measurements and 6,971 of them with distance measurements. The typical error on metallicity is around 0.2\,dex, and the typical systematic radial velocity uncertainty is from 5 to 21 \kms, which depends on the number of spectra used for deriving the radial velocity for a star. 

By cross matching with {\it Gaia} EDR3 \citep{GaiaCollaboration2021} using a search radius of one arcsec, we get the proper motions for 8,022 stars in our sample, from which the number of stars with {\tt ruwe} $<$ 1.4 \citep{Lindegren2018, Lindegren2021, Fabricius2021} is 7,795. In order to study the detailed properties of the Galactic halo, 
we then select sources with [Fe/H] $\leq -1.0$\,dex, and
$|Z| \ge$ 3 \,kpc (which excludes the potential contaminators from the Galactic disk). The final number of stars with metallicity and distance measurements is 5,252.

 \begin{figure*}
  \centering
  \includegraphics[scale=0.6, angle=0]{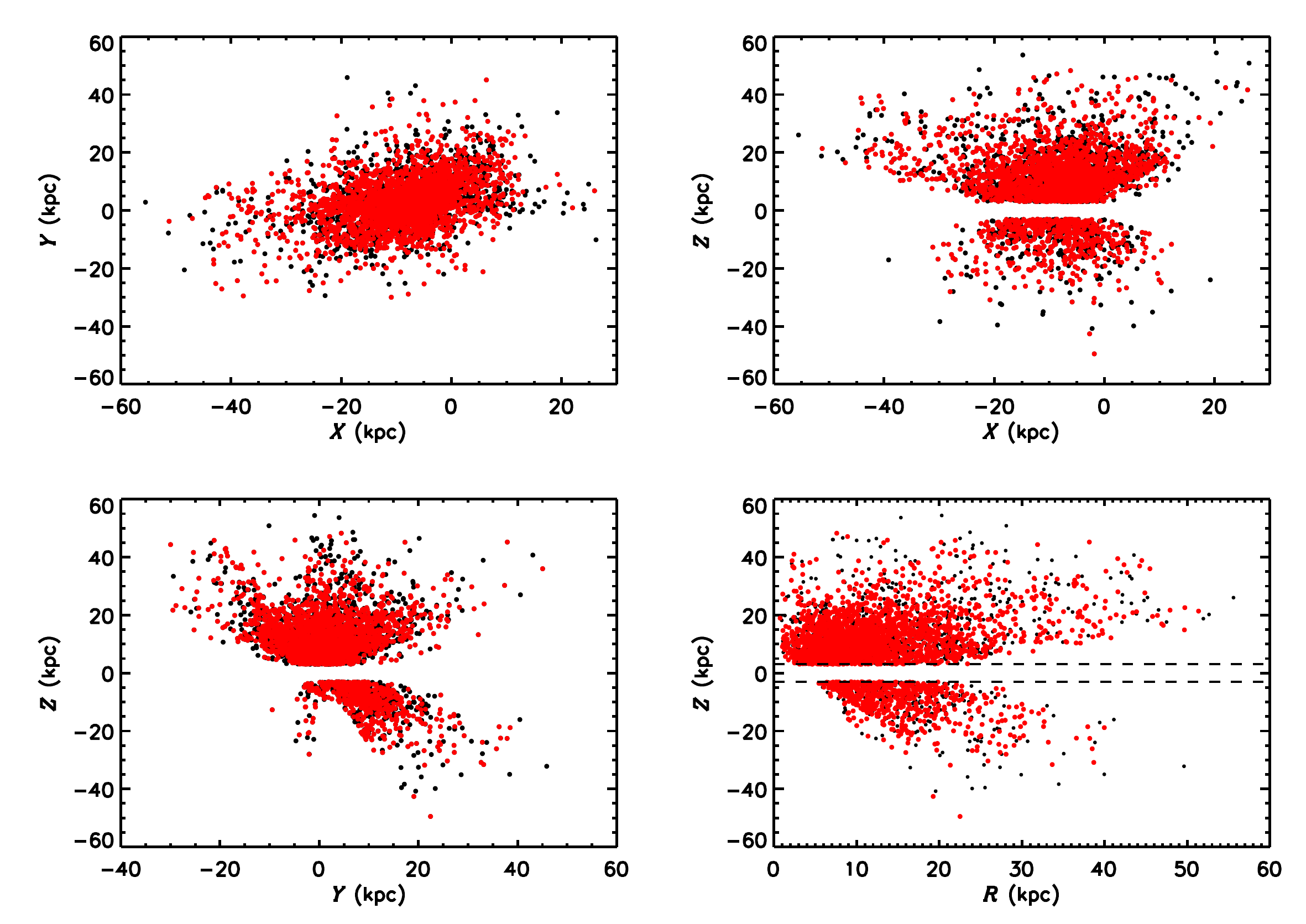}
  \caption{ Spatial distributions of our final sample in $X-Y, X-Z, Y-Z$ and $R-Z$ panels where $R = \sqrt{X^2+Y^2}$ and $X, Y, Z$ are in Galactocentric Cartesian coordinates. Black dots represent the whole final sample (4,365) and red dots represent the stars with systemic radial velocities measurements (2,954). The dashed lines in the lower right panel mark |Z| = 3 kpc, which we use to excise stars located in the disk. }
   \label{fig1}
\end{figure*}

\begin{figure*}
\centering
\includegraphics[width=16cm]{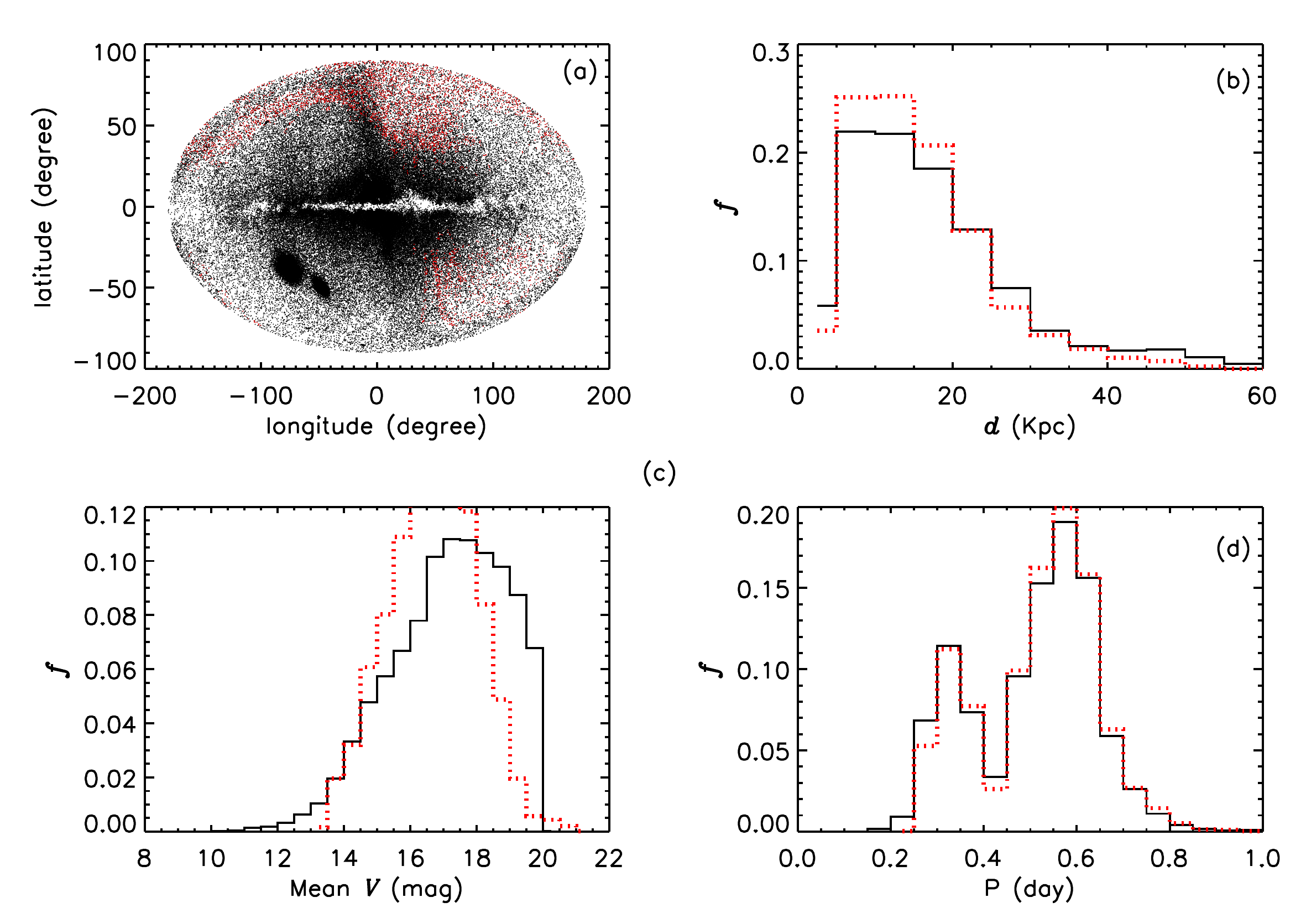}
\caption{ Basic properties of our photometric (black dots/lines, 206,664) and final spectroscopic (red dots/lines, 4,365) RRL samples. Panel (a) shows the spatial distribution in Galactic coordinates, Panel (b) the normalized distribution of distances collected from the literature, Panel (c) the normalized distribution of mean $V$ band magnitudes, and Panel (d) the normalized distribution of periods.} 
\label{fig2}
\end{figure*}

\subsection{Coordinate systems}
\label{transformation}

We now have the following quantities for each star: right ascension and
declination, proper motions and heliocentric radial velocity. Next, we transform the observables
to spherical polar coordinates in the Galactic rest-frame.  To account
for measurement errors, we propagate the errors of the observables (assuming Gaussian distributions) to our Galactocentric spherical coordinates by Monte-Carlo simulations. 

We use a right-handed Galactocentric Cartesian coordinate system
$( X , Y , Z )$
and a Galactocentric spherical coordinate system $( r , \theta , \phi)$.
In the
Galactocentric Cartesian coordinate system, $X$ points from the Galactic centre to the direction away from the Sun, $Y$ points in the direction such that the local standard of rest moves in the positive $Y$ direction and $Z$ points toward
the North Galactic Pole.
In the Galactocentric spherical coordinate
system, $r$ is the Galactocentric distance, the polar angle $\theta$ increases from 0 to $\pi$
from the North Galactic Pole to the South Galactic Pole and the azimuthal angle $\phi$ is between the
direction from the Galactic centre toward the Sun and the direction
to the projected position of the star.
The three Galactocentric spherical 
velocity components are represented by
($v_r, v_\theta, v_\phi$).
We use the solar motion values $v_{\odot} = (7.01,
252, 4.95)$ \kms and distance of the Sun to the Galactic centre $R_0 = 8.34$ kpc, which are taken from \citet{Huang2015}, \citet{Reid2004} and \citet{Reid2014}. 

Similarly with \citet{Wang2022} (\citetalias{Wang2022}), we exclude the
stars with large uncertainty on any one component in velocity ($\sigma > $100 \kms) and the stars with very large total velocity, namely, $\sqrt{v_r^2+v_\theta^2+ v_\phi^2} > 400$ \kms. This cut gets rid of 32 stars in total. The stars removed by these cuts are uniformly distributed in distance and metallicity.

\subsection{Removal of Sagittarius stream members} 
\label{subsec:Sagittarius_removal}
 In the past decades, many substructures have been uncovered. Examples are the Helmi streams \citep{Helmi1999Nature}, Sagittarius stream \citep{Ivezic2000, Yanny2000}, Virgo overdensity \citep{Vivas2004}, GES \citep{Belokurov2018, Helmi2018}, Sequoia \citep{Myeong2019}, and Thamnos \citep{Matsuno2019}. On account of the low number density of RRL, we ignore the effect of small substructures (for more details about these small substructures, please refer to \citetalias{Wang2022}). At the same time, one of our goals is to explore the properties of GES, so here we only exclude the members of the Sagittarius stream, which appears to be one of the most significant substructures of the halo of our Galaxy. 

To remove the members of the Sagittarius stream, we convert all stars to the Sagittarius coordinate system, using the transform matrix given by \citet{Belokurov2014}. We then use the positions
of the stream defined by \citet{Hernitschek2017} to remove stars in $\Lambda_\odot-d$ plane. Here only stars within 10$^{\circ}$ of the Sagittarius stream plane are considered. Following equations 2 and 3 of \citet{Lancaster2019},  we remove 801 potential member stars of the Sagittarius stream. This filtering process reduces our sample size to 4,365 stars.
The spatial distribution of the final sample is shown in Figure\,\ref{fig1}. In the next section, we use this sample (black dots) to analyze the distribution in metallicity. When we analyze the properties of the kinematics, we need the systematic radial velocity, so the sample size reduces to 2,954 stars (red dots).

 \begin{figure}
  \centering
  \includegraphics[scale=0.25, angle=0]{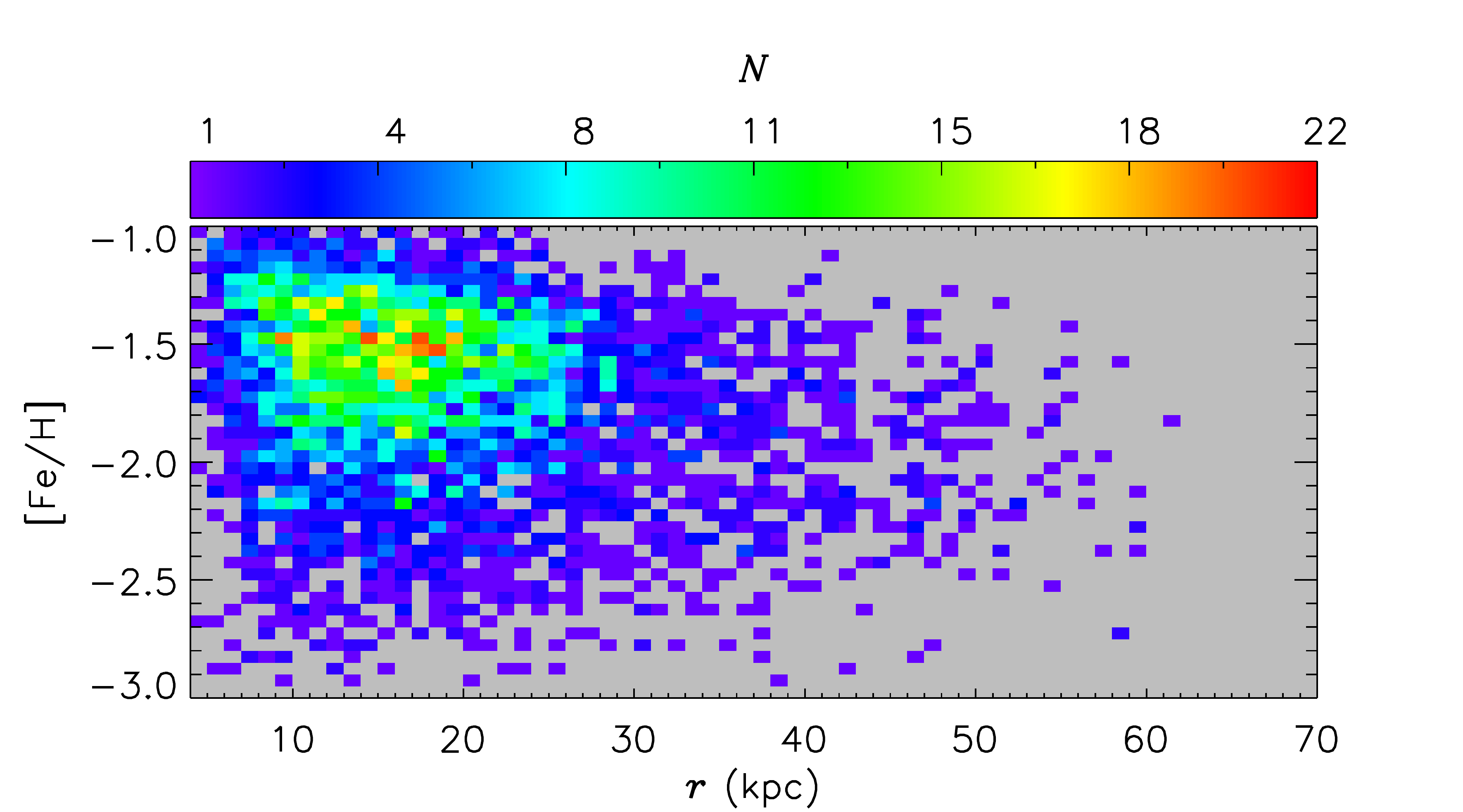}
  \caption{ Density distribution of metallicity for RRL in the Galactic halo.}
   \label{fig3}
\end{figure}

\begin{figure*}
 \centering
  \includegraphics[scale=0.6, angle=0]{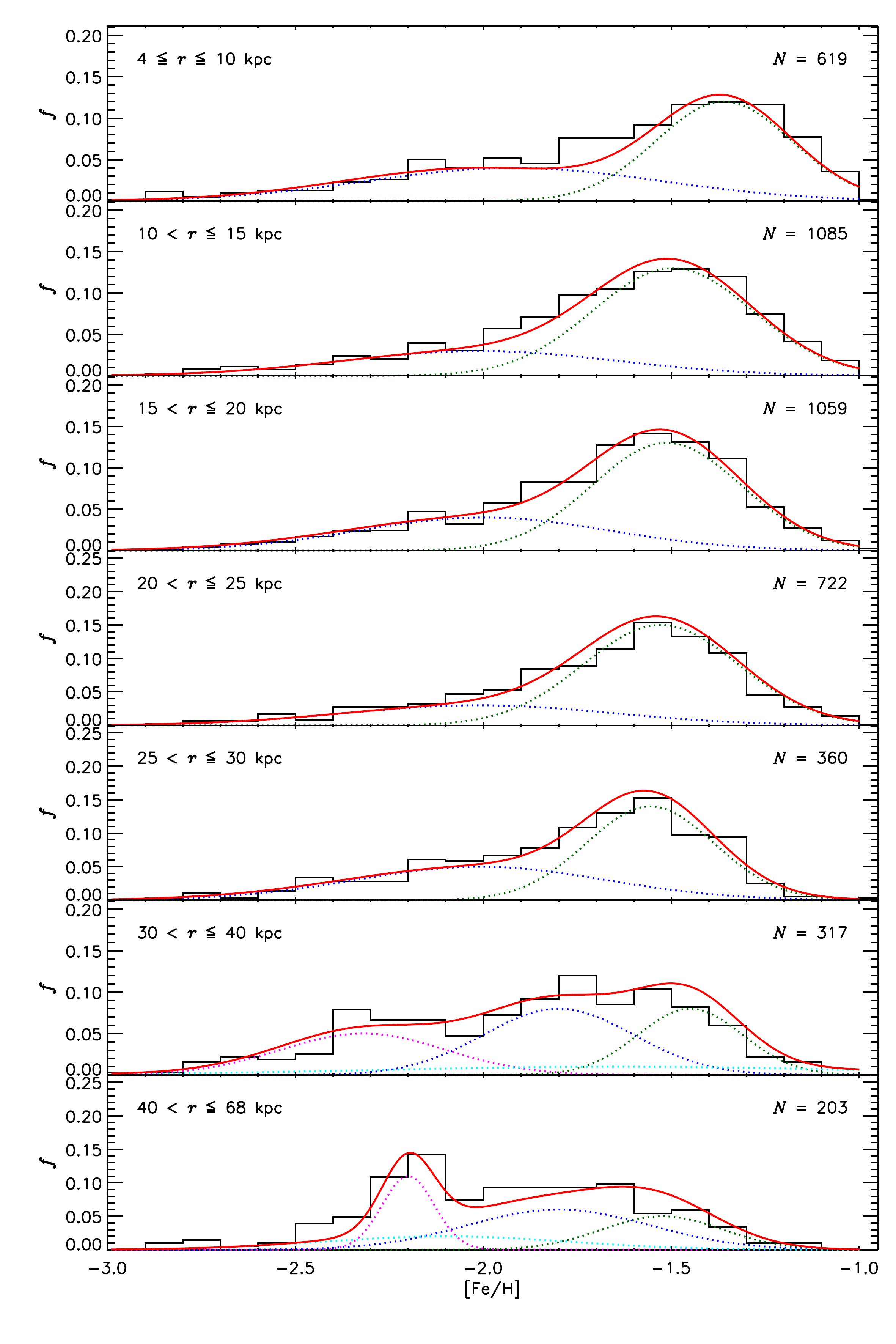}
  \caption{MDFs of seven radial bins for the stellar halo. The radial distance range is shown in the upper-left corner and the number in each bin is shown in the upper-right corner. The  dotted lines represent the contributions from individual assumed components and the red solid line represents the total summed results.}
   \label{fig5}
\end{figure*}

\begin{figure*}
 \centering
\includegraphics[scale=0.4,angle=0]{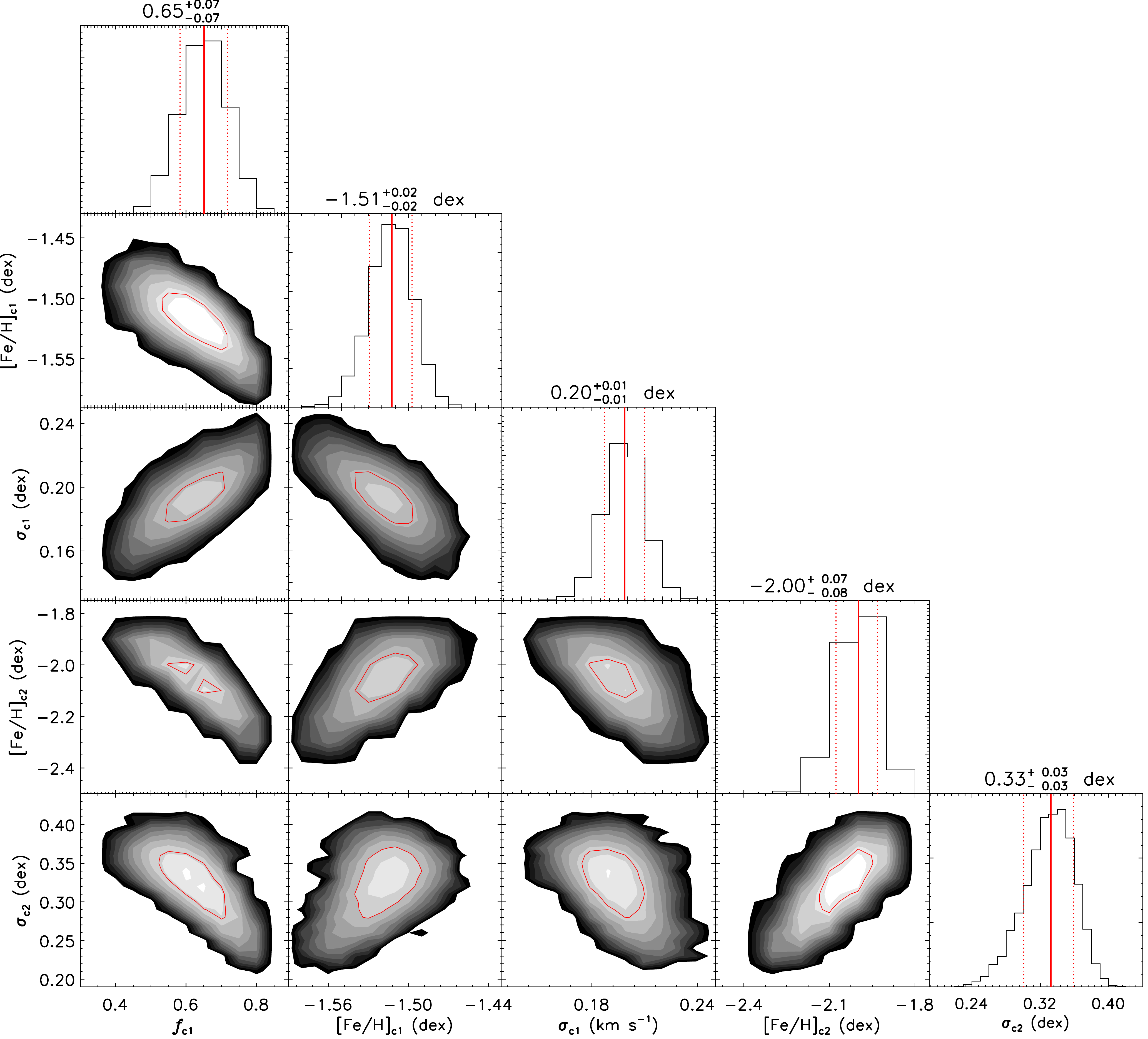}
\caption{
Two and one dimensional projections of the marginalized posterior probability distribution of the five model parameters (see details in Section\,\ref{subsec:metallicity}) obtained by the Bayesian MCMC technique applied to the radial bin $15 < r \leq 20$\,kpc.
The red contour of each two dimensional distribution denotes the $1\sigma$ confidence level; and the red solid and dotted vertical lines of each one dimensional histogram denote the 50 per cent and 68 per cent probability intervals, respectively, for each parameter. The yielded values and uncertainties for each parameter are also shown on the top of each column.}
\label{fig4}
\end{figure*}

  \begin{figure}
  \centering
  \includegraphics[scale=0.37, angle=0]{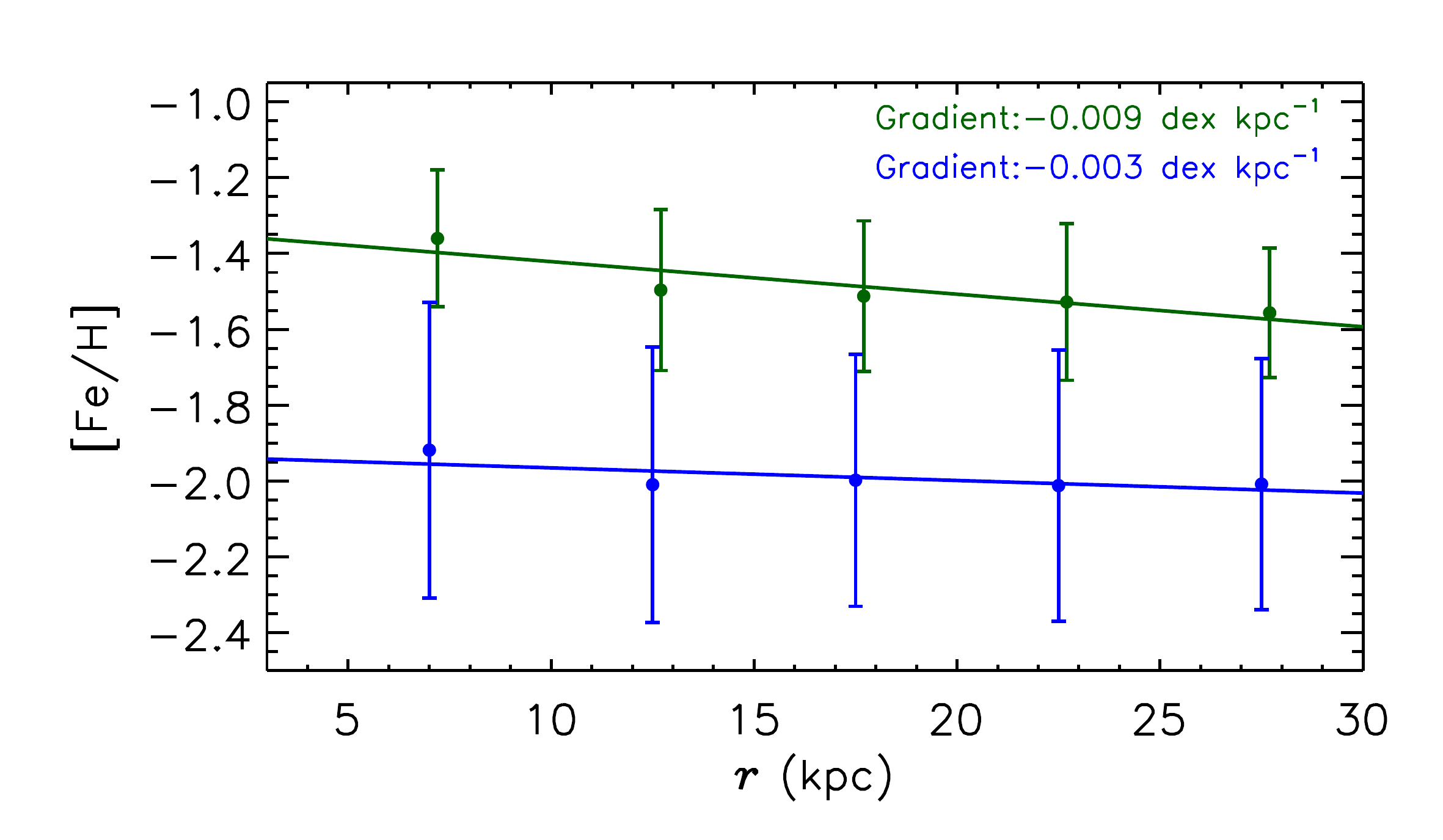}
  \caption{ Gradient of metallicity for two components, namely GES (green) and a metal-poor component (blue) in the inner halo. Each point marks the mean metallicity of the fitted Gaussians and the error bars show the corresponding metallicity dispersion $\sigma$ (see Section\,\ref{subsec:metallicity}) as a function the median distance $r$ within a bin. For clarity the markers for the GES components are shifted by 0.2 kpc along the $x$-axis. }
   \label{fig_gradient}
\end{figure}

\section{Results}

In this section, we use the chemical abundance distribution, anisotropy parameter $\beta$ and velocity distribution of RRL to explore the properties of the Galactic halo.

\subsection {Metallicity} 
 \label{subsec:metallicity} 
   
   The absolute magnitudes of RRL are only very weakly dependent on the metallicity, so the photometric sample of RRL  do not suffer from metallicity selection effects. By the aforementioned process of clipping, we obtain 4,365 stars as our final spectroscopic sample to study the Galactic halo. We now compare the properties of our photometric and spectroscopic samples. Figure\,\ref{fig2} plots the distribution of the stars in Galactic coordinates and normalized histogram of distance from the Sun, mean $V$ band magnitude and period (Panels (a)$-$(d), respectively). For the sample observed by {\it Gaia}, we convert the $G$ band magnitudes to $V$ band magnitudes by using equation 2 of \citet{Clementini2019}. In Panel (c) of Figure\,\ref{fig2}, we only show the distribution of stars brighter than 20 mag (143,928) for the photometric sample, since the magnitude of the spectroscopic sample (SDSS and LAMOST) is brighter than 20 mag. Panel (d) of Figure \ref{fig2} shows the period distribution of the photometric and spectroscopic samples are almost the same. Owing to the apparent period-luminosity relation of RRL, the luminosity distribution of the spectroscopic sample is similar to that of the photometric sample. By the above comparisons, we generally think that the final spectroscopic sample does not suffer significant selection effects.  Our RRL spectroscopic sample is therefore suitable for studying the density and MDF of the Galactic halo.

   Figure\,\ref{fig3} shows the density distribution of metallicities of our final RRL sample (4,365 stars). We find there exists an obvious metal rich component with a mean metallicity of $-1.5$\,dex, largely ranging from $-1.1$ to $-1.8$\,dex. This metal-rich component is mostly distributed in the inner part with $r$ smaller than 30\,kpc. The metallicity density of this metal rich component is apparently larger than other parts of the Galactic halo, which indicates they may have different origins.  For $r > 30$ kpc,  the distribution of metallicity is very broad which is expected for a stellar halo formed from the debris of disrupted satellites. 
   According to the density profile of halo stars selected from SDSS, \citet{Carollo2007} put forward a dual component model of the Galactic halo (inner halo and outer halo). The metallicity peaks at $-1.6\,$dex for the inner halo and $-2.2\,$dex for the outer halo. \citet{Belokurov2018} found a significant halo component with radial orbits ($\beta \sim  0.9$) in the inner halo, and demonstrated that such stars are the relic from the merger of a large metal-rich dwarf galaxy with mass $>10^{10} \, \mathrm{M}_\odot$.  For the metal-rich component probed by our RRL sample, we find a sharp decline in the number density at $r \sim$ 30 kpc. For $r <$ 30 kpc, the inner sample of RRL is mainly dominated by the metal rich component, but for $r > $ 30 kpc, the metallicity distribution is more broad. A break radius $r_b$ = 27.8 kpc was also found by \citet{Sesar2011} from the density profile of RRL, after which the slope of the profile becomes steeper. Our break radius $r_b$ = 30 kpc is close to the long-established break radius of the classical stellar halo.
   
  We further divide the RRL into seven radial bins according to their Galactocentric distances from 4 to 68 kpc in steps of 5 kpc, except the first ($4 \leq r < 10\,$kpc) and the last two ( $30 \leq r < 40\,$kpc, $40 \leq r < 68\,$kpc) bins. We set the bin size of metallicity equal to 0.1\,dex since the measurement error of metallicity is around 0.2\,dex. This binsize also ensures a large enough number of stars, typically $>30$, in each metallicity bin. Figure\,\ref{fig5} shows the MDF for each radial bin. We find that the MDFs of the stellar halo show significant differences between $ 4 \leq r < 30$ kpc (inner halo)  and  $ 30 \leq r < 68$ kpc  (outer halo). For the inner halo, the metal-rich stars completely dominate the MDF and the peak value of metallicity becomes poorer with increasing $r$; but for the outer halo, the metallicity distribution for RRL transitions to a much broader composition of several weak peaks. 

The distinct metallicity density map and MDFs for our RRL sample at $r <$ 30 kpc and $r >$ 30 kpc are best explained with a dual (at least two) component halo model as proposed by previous investigations using different halo tracers \citep[e.g.][]{Chiba2001, Carollo2007, Carollo2010, Belokurov2018, Helmi2018}. 
 \citet{Carollo2007, Carollo2010} explain the radial gradient of metallicity for the inner halo as caused mainly by radial mergers of massive clumps followed by a stage of adiabatic compression. \citet{Belokurov2018} put forward that the inner halo contains a large number of metal-rich stars which are the relic of a major merger event involving a metal-rich dwarf galaxy. Our results also support that the inner halo is mainly dominated by metal rich stars, the likely relics of this major merger with a metal-rich dwarf galaxy. At the same time, the long metal-poor tail of the RRL inner halo MDF is consistent with the build up over-time of the stellar halo through multiple minor mergers \citep{White1978}. The outer halo RRL are characterized by a very broad MDF composed of several weak peaks. Multiple weak peaks are likely indications of multiple minor mergers gradually building up the stellar halo \citep{Myeong2018.475}.

In order to explore the chemical properties of the GES, we use two Gaussian components (one accounts for the GES and the other for the remaining halo component) to fit the MDF for the inner halo, and multi-Gaussian components for the outer halo. 
For each radial bin in the inner halo, the metallicities of GES stars are assumed to follow a Gaussian distribution with a mean $\rm[Fe/H]_{\rm c1}$ and a dispersion $\sigma_{\rm c1}$. 
The number fraction of GES stars is represented by $f_{\rm c1}$. The metallicity distribution of the remaining halo component is also represented by a Gaussian distribution characterized by mean $\rm[Fe/H]_{\rm c2}$ and dispersion $\sigma_{\rm c2}$. In this way, for the model parameters $\Theta =$\,\{$\rm[Fe/H]_{\rm c1}$,\,$\sigma_{\rm c1}$,\,$f_{\rm c1}$,\,$\rm[Fe/H]_{\rm c2}$,\,$\sigma_{\rm c2}$\}, the likelihood of observing the $i$th sample star with metallicity ${\rm [Fe/H]}_{i}$ is given by,

\begin{equation}
L_{i} (\rm[Fe/H]_{i}|\Theta) = \it{f}_{\rm c1}\it{P}_{\rm c1} ({\rm [Fe/H]}_{i}) + (\rm{1} - \it{f}_{\rm c1})\it{P}_{\rm c2}({\rm [Fe/H]}_{i})\text{.} 
\label{eq1}
\end{equation} 

The probability $P_{\rm c1} ({\rm [Fe/H]}_{i})$ of a GES star with a measured metallicity [Fe/H]$_{i}$ can be calculated from,

\begin{equation}
P_{\rm c1} ({\rm [Fe/H]}_{i}) = \frac{1}{\sqrt{2\pi}\sigma_{\rm c1}}\exp{\left[-\frac{1}{2}\left(\frac{{\rm [Fe/H]}_{i}-\rm[Fe/H]_{\rm c1}}{\sigma_{\rm c1}}\right)^2\right]}\text{.} 
\label{eq2}
\end{equation}
       
  The probability $P_{\rm c2}({\rm [Fe/H]_{i}})$ of a star with measured metallicity $\rm{[Fe/H]}_{i}$ belonging to the remaining halo component can be obtained from,
  
\begin{equation}
P_{\rm c2}({\rm [Fe/H]}_{i}) = \frac{1}{\sqrt{2\pi}\sigma_{\rm c2}}\exp{\left [-\frac{1}{2}\left (\frac{{\rm [Fe/H]}_{i}-\rm[Fe/H]_{\rm c2}}{\sigma_{\rm c2}}\right )^2\right ]}\text{.} 
\end{equation}

The likelihood of a specific radial bin $j$ is calculated by multiplying the function of equation \ref{eq1} for a total of $N_j$ stars located within bin $j$,

\begin{equation}
\centering
L = \prod_{i=1}^{N_j} L_i\text{.} 
\end{equation}

The posterior distribution of the model parameters is given by,

\begin{equation}
p (\Theta|{\boldsymbol {O}}) \propto L({\boldsymbol {O}}|\Theta)I(\Theta)\text{,}
\end{equation}

where \textbf{$O$} represents the observables, i.e. \,[Fe/H]$_{i}$
and $I (\Theta) $ encapsulates the priors of the model parameters.
Table \ref{prior} presents the details of the priors of model parameters for individual radial bins. In this study, the Bayesian Markov chain Monte Carlo (MCMC) technique is used to obtain the posterior probability distributions of the model parameters. As an example, Figure\,\ref{fig4} shows the posterior probability distributions yielded by the Bayesian MCMC technique of the five free parameters we assumed in modeling the MDF at radial bin $15 < r < 20$\,kpc.
The best-fit values and uncertainties of the model parameters are then properly delivered by the marginalized posterior probability distributions shown in Figure \ref{fig4}.

For the outer-halo, we use two, three or four Gaussian components to fit the MDF using the same method described above. The values of the Bayesian Information Criterion \citep[BIC,][]{Schwarz1978} are calculated to evaluate the different models and are presented in Table\, \ref{bic}. The best model is four Gaussian components to fit the MDF for radial bin $30 < r \leq 68$\,kpc, which means the outer halo likely experienced several minor mergers. 
All the best-fit results are shown in Figure\,\ref{fig5} and Table\, \ref{bvalue}. 

\begin{table*}
 \caption{Priors for the parameters of the metallicity distribution model.}
 \centering
 \begin{threeparttable}
\begin{tabular}{l|ll|ll|ll}
\hline
\hline
Region                             & \multicolumn{2}{c|}{4 $< r \leq$ 30 kpc}              &\multicolumn{2}{c}{30 $< r \leq 68$ kpc} & 30 $< r \leq 40$ kpc & 40 $< r \leq$ 68 kpc\\
\hline
{Model }                              & Two Gaussians & Three Gaussians  &Two Gaussians & Three Gaussians & Four Gaussians & Four Gaussians\\
{[Fe/H]$_{\rm c1}$ (dex)}  & $[-1.7, -1.0]$& $[-1.5, -1.0]$ &$[-1.0, -2.0] $& $[-1.0, -1.5]$& $[-1.0, -1.5]$ & $[-1.0, -1.7]$\\
$\sigma_{c1}$ (dex)       & $[0.0, 0.6]$ & $[0.0, 0.2]$&$[0.0, 1.0]$ & $[0.0, 0.3]$&$[0.0, 0.2]$ &$[0.0, 0.2]$\\
$f_{c1}$                          & $[0.0, 1.0]$& $[0.1, 0.5]$& $[0.2, 0.8]$& $[0.1, 0.8]$& $[0.1, 0.5]$ & $[0.1, 0.5]$\\
{[Fe/H]$_{\rm c2}$ (dex)}  & $[-3.0, -1.7]$& $[-1.5, -2.0]$& $[-2.0, -3.0]$& $[-2.0, -3.0]$&$[-1.7, -2.0]$ &$[-1.7, -2.0]$\\
$\sigma_{c2}$                &$[0.0, 1.0]$ & $[0.0, 0.5]$& $[0.2, 0.6]$& $[0.0, 0.2]$&$[0.0, 0.3]$ &$[0.0, 0.3]$\\
$f_{c2}$                          &--  & $[0.3, 0.8]$       &--    & $[0.1, 0.5]$    &$[0.1, 0.5]$ &$[0.1, 0.5]$\\
{[Fe/H]$_{\rm c3}$ (dex)} &-   &$[-3.0, -2.0]$ &-- & $[-1.5, -2.0]$ &$[-2.2, -3.0]$ &$[-2.2, -3.0]$\\
$\sigma_{c3}$  &-- & $[0.2, 1.0]$   &-- & $[0.0, 0.2]$    &$[0.0, 0.3]$ &$[0.0, 0.3]$\\
$f_{c3}$                         &-- &-- &--&--&$[0.1, 0.5]$ &$[0.1, 0.5]$\\
{[Fe/H]$_{\rm c4}$ (dex)} &-- &-- &--&--&$[-1.9, -2.2]$&$[-1.7, -2.5]$\\
$\sigma_{c4}$                &-- &-- &--&-- &$[0.5, 1.0]$  &$[0.1, 0.8]$\\ 
\hline
\end{tabular}
\begin{tablenotes}
        \footnotesize
       \item Note: All parameters are uniform distributed between the two limits shown in the square brackets.
      \end{tablenotes}
\end{threeparttable}
\label{prior}
\end{table*}

 \begin{table}
\centering
\caption{Values of Bayesian Information Criterion of different models for the outer-halo.}
\begin{tabular}{ccc}
\hline
\hline
Model&$30 < r \leq 40$ kpc &$40 < r \leq 68$ kpc\\
\hline
Two Gaussian component&292.16  &157.99    \\
Three Gaussian component& 306.00 &206.53    \\
Four Gaussian component& 246.06 &128.87    \\
\hline
\end{tabular}
 \label{bic}
\end{table}

\begin{table*}
\caption{Fitting results for our seven radial bins.}
\centering
\begin{tabular}{c|c|c|c|c|c|c|c}
\hline
\hline
Radial bins (kpc) &[4, 10]&[10, 15]&[15, 20]&[20, 25]&[25, 30]&[30, 40]&[40, 68]\\
\hline
[Fe/H]$_{\rm c1}$ (dex)&$-1.36^{+0.02}_{-0.02}$&$-1.50^{+0.02}_{-0.02}$&$-1.51^{+0.02}_{-0.02}$&$-1.53^{+0.02}_{-0.02}$&$-1.56^{+0.02}_{-0.03}$&$-1.50^{+0.17}_{-0.10}$&$-1.52^{+0.11}_{-0.10}$\\
$\sigma_{c1}$ (dex)    & $0.18^{+0.02}_{-0.01}$ &$0.21^{+0.01}_{-0.01}$&$0.20^{+0.01}_{-0.01}$&$0.21^{+0.02}_{-0.01}$&$0.17^{+0.02}_{-0.02}$&$0.15^{+0.04}_{-0.06}$ &$0.16^{+0.03}_{-0.05}$\\
$f_{c1}$                       &$0.51^{+0.06}_{-0.05}$ &$0.68^{+0.05}_{-0.06}$&$0.65^{+0.07}_{-0.07}$&$0.67^{+0.08}_{-0.08}$&$0.53^{+0.09}_{-0.08}$&$0.26^{+0.13}_{-0.10}$&$0.11^{+0.09}_{-0.07}$\\
$\rm{[Fe/H]}_{\rm c2}$ (dex)&$-1.92^{+0.05}_{-0.05}$&$-2.01^{+0.06}_{-0.07}$&$-2.00^{+0.07}_{-0.08}$&$-2.01^{+0.08}_{-0.10}$&$-2.01^{+0.06}_{-0.08}$&$-1.80^{+0.00}_{-0.00}$&$-1.80^{+0.00}_{-0.00}$\\
$\sigma_{c2}$ (dex)   &$0.39^{+0.02}_{-0.02}$&$0.36^{+0.02}_{-0.03}$&$0.33^{+0.03}_{-0.03}$&$0.36^{+0.03}_{-0.03}$&$0.33^{+0.03}_{-0.03}$&$0.20^{+0.07}_{-0.09}$&$0.24^{+0.04}_{-0.09}$\\
 $f_{c2}$   &--&--&--&--&--&$0.32^{+0.12}_{-0.13}$  &$0.30^{+0.12}_{-0.13}$\\
$\rm{[Fe/H]}_{\rm c3}$ (dex)&-- &-- &-- & --&--&$-2.30^{+0.06}_{-0.06}$&$-2.20^{+0.02}_{-0.02}$\\
 $\sigma_{c3}$ (dex) &-- &-- &-- & --&--&$0.21^{+0.06}_{-0.09}$&$0.07^{+0.09}_{-0.07}$\\
 $f_{c3}$    &-- &-- &-- & --&--&$0.29^{+0.11}_{-0.12}$&$0.13^{+0.15}_{-0.03}$\\
 { [Fe/H]$_{\rm c4}$} (dex)&-- &-- &-- & --&--&$-1.60^{+0.00}_{-0.00}$&$-2.10^{+0.00}_{-0.00}$\\
  $\sigma_{c4}$ (dex)&-- &-- &-- & --&--&$0.65^{+0.20}_{-0.11}$&$0.33^{+0.07}_{-0.04}$\\
\hline
\end{tabular}
\label{bvalue}
\end{table*}
 
\begin{figure}
  \centering
  \includegraphics[scale=0.4, angle=0]{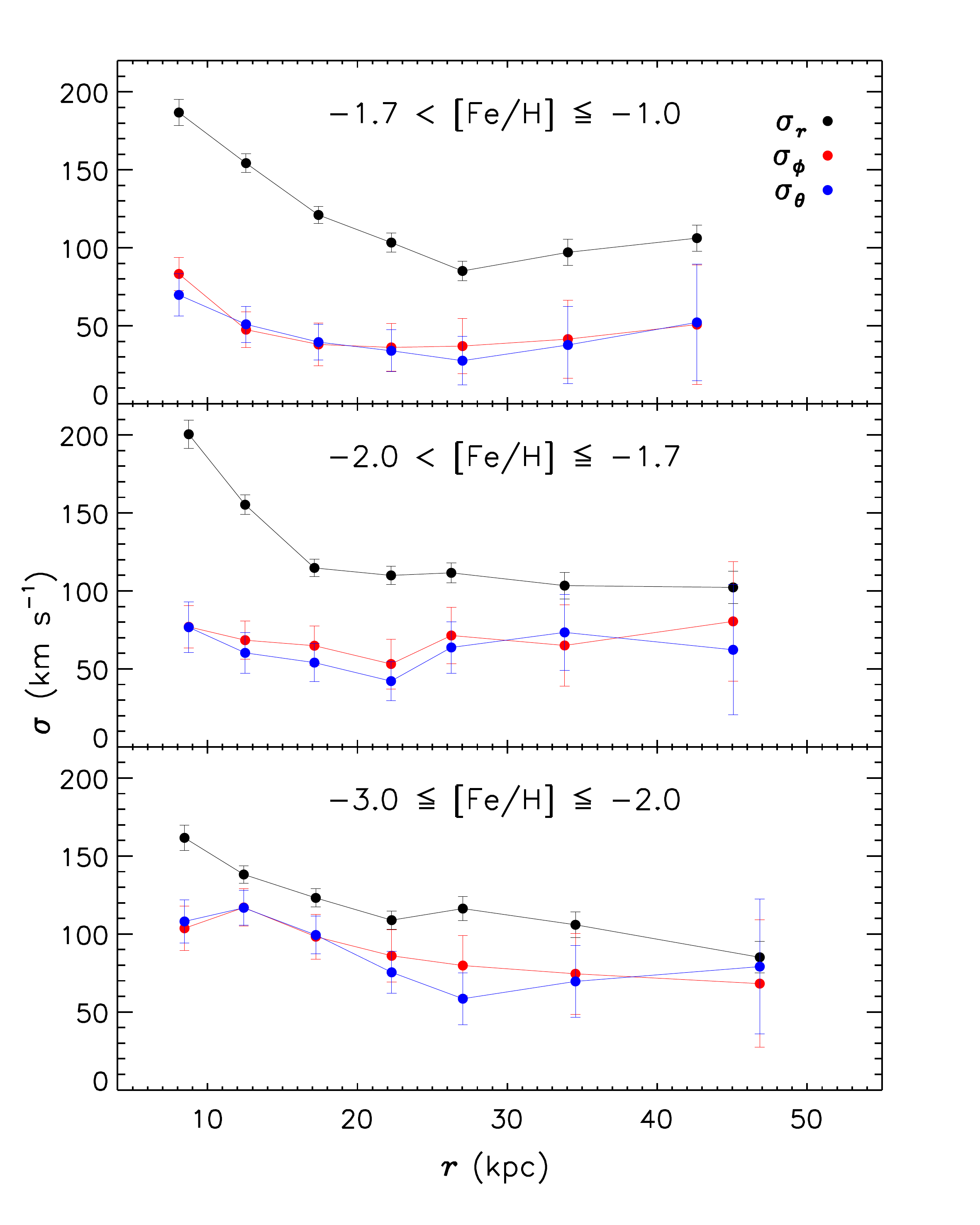}
  \caption{ Velocity dispersion for three different metallicity populations ($- 1.7 < \rm{[Fe/H]} \leq -1.0$, $- 2.0 < \rm{[Fe/H]} \leq -1.7$, $- 3.0 \leq \rm{[Fe/H]} \leq -2.0$) as a function of Galactocentric distance $r$. }
   \label{fig6}
\end{figure}

 \begin{figure}
   \centering
  \includegraphics[scale=0.35, angle=0]{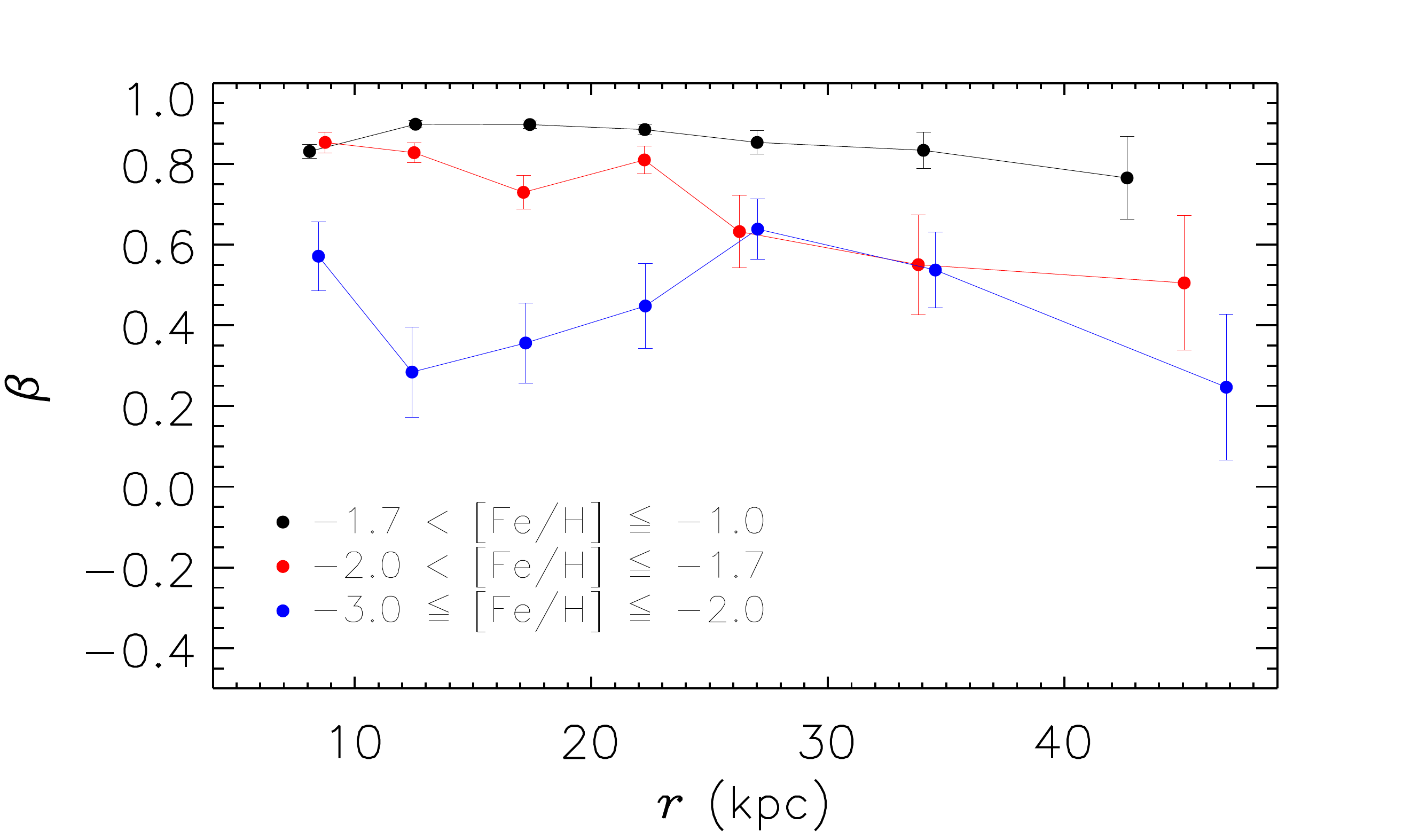}
    \caption{ Anisotropy parameter $\beta$ as a function of $r$ for our three metallicity bins.}
   \label{fig7}
\end{figure}
   
 \begin{figure}
   \centering
  \includegraphics[scale=0.25, angle=0]{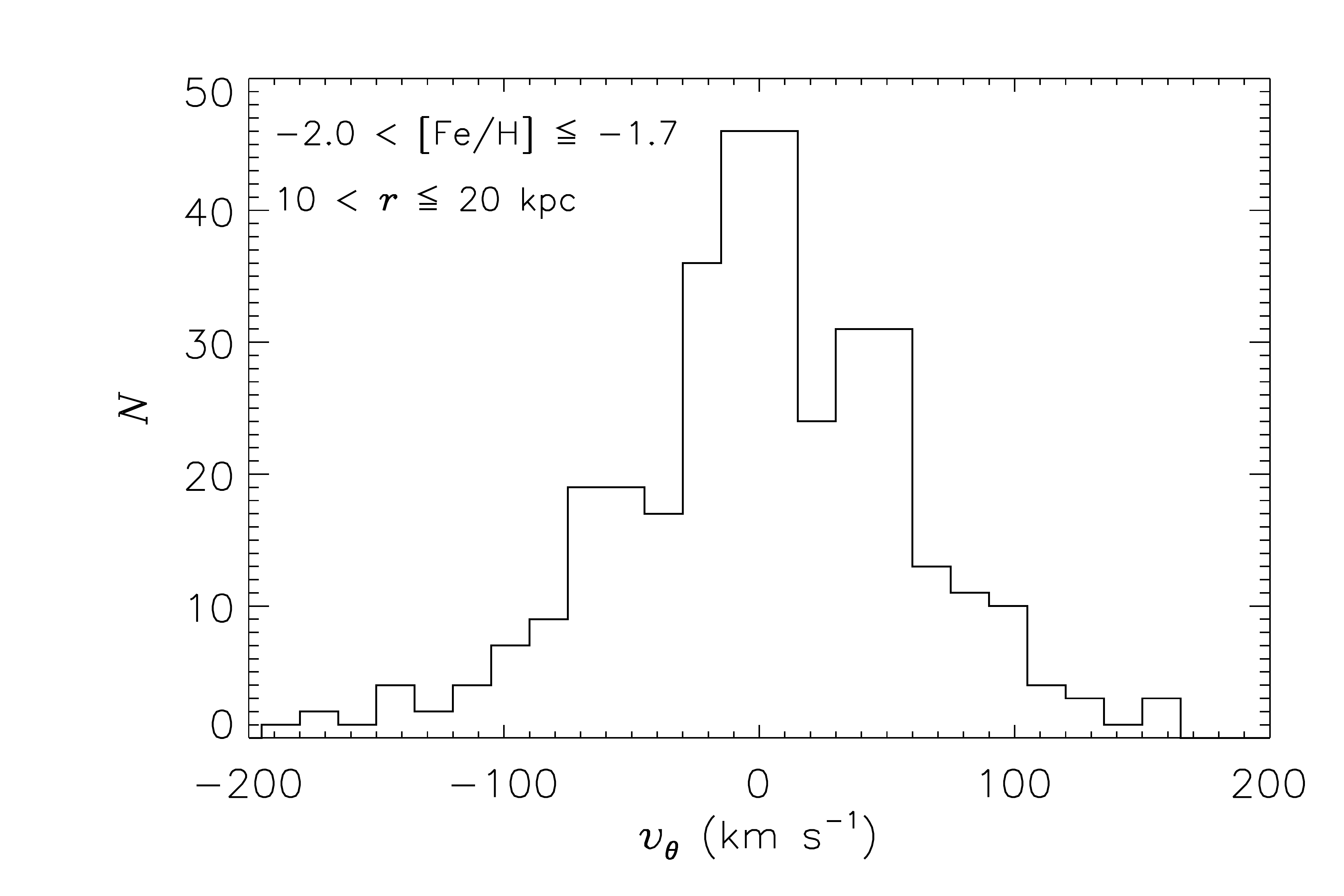}
    \caption{ Distribution of $v_{\theta}$ for metal-medium population located between 10 and 20 kpc.}
   \label{subs}
\end{figure}

 \begin{figure*}
  \centering
  \includegraphics[scale=0.35, angle=0]{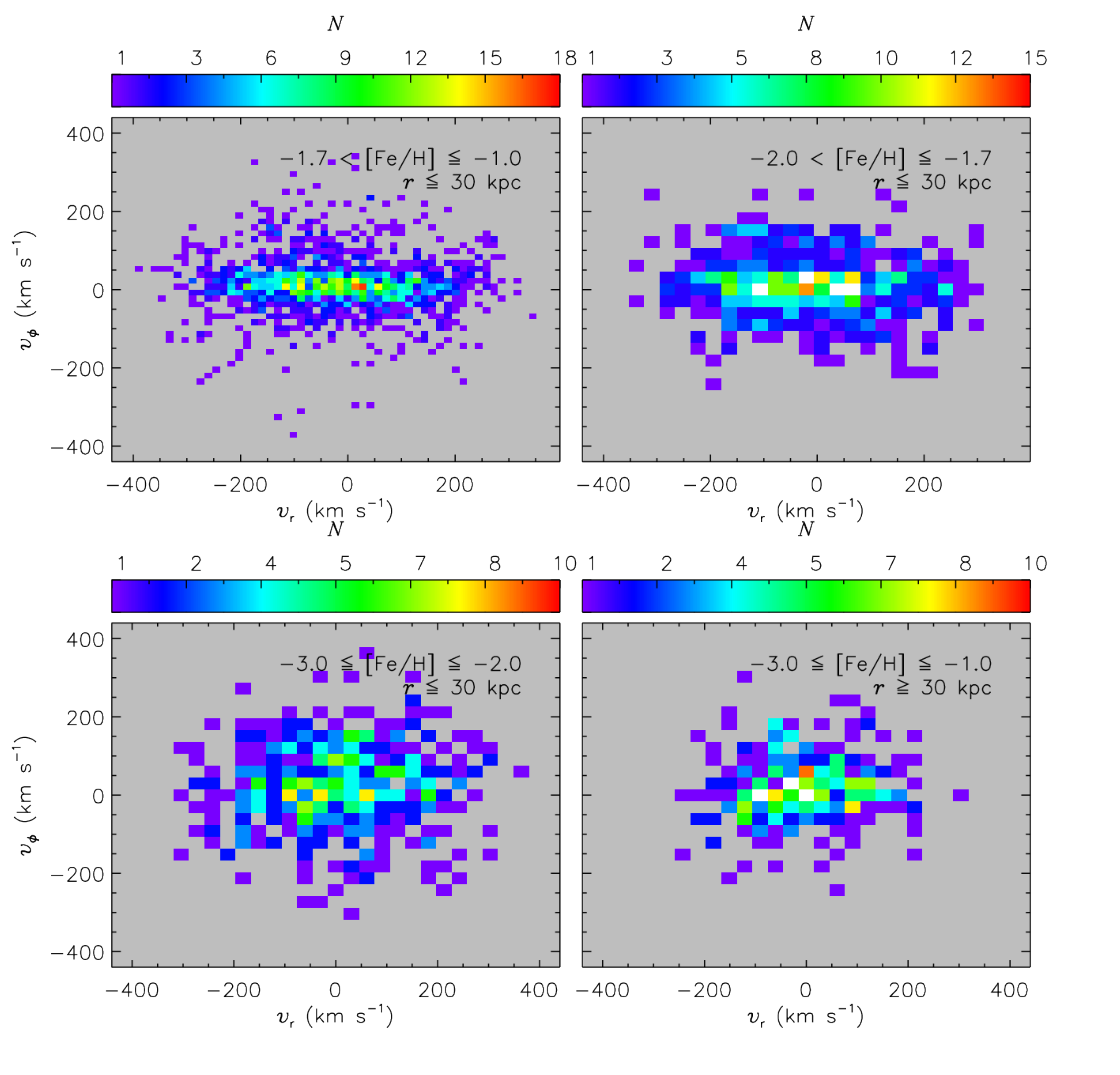}
    \caption{ Radial $v_{r}$ and rotational  $v_{\phi}$ velocity distributions for  three populations bases on metallicity at $r \leq 30$ kpc (inner halo) and for the whole population in outer halo ($r \geq 30$ kpc). The color bar indicates the number of stars $N$.}
   \label{fig:fig7}
\end{figure*}

  \begin{figure*}
  \centering
  \includegraphics[scale=0.45, angle=0]{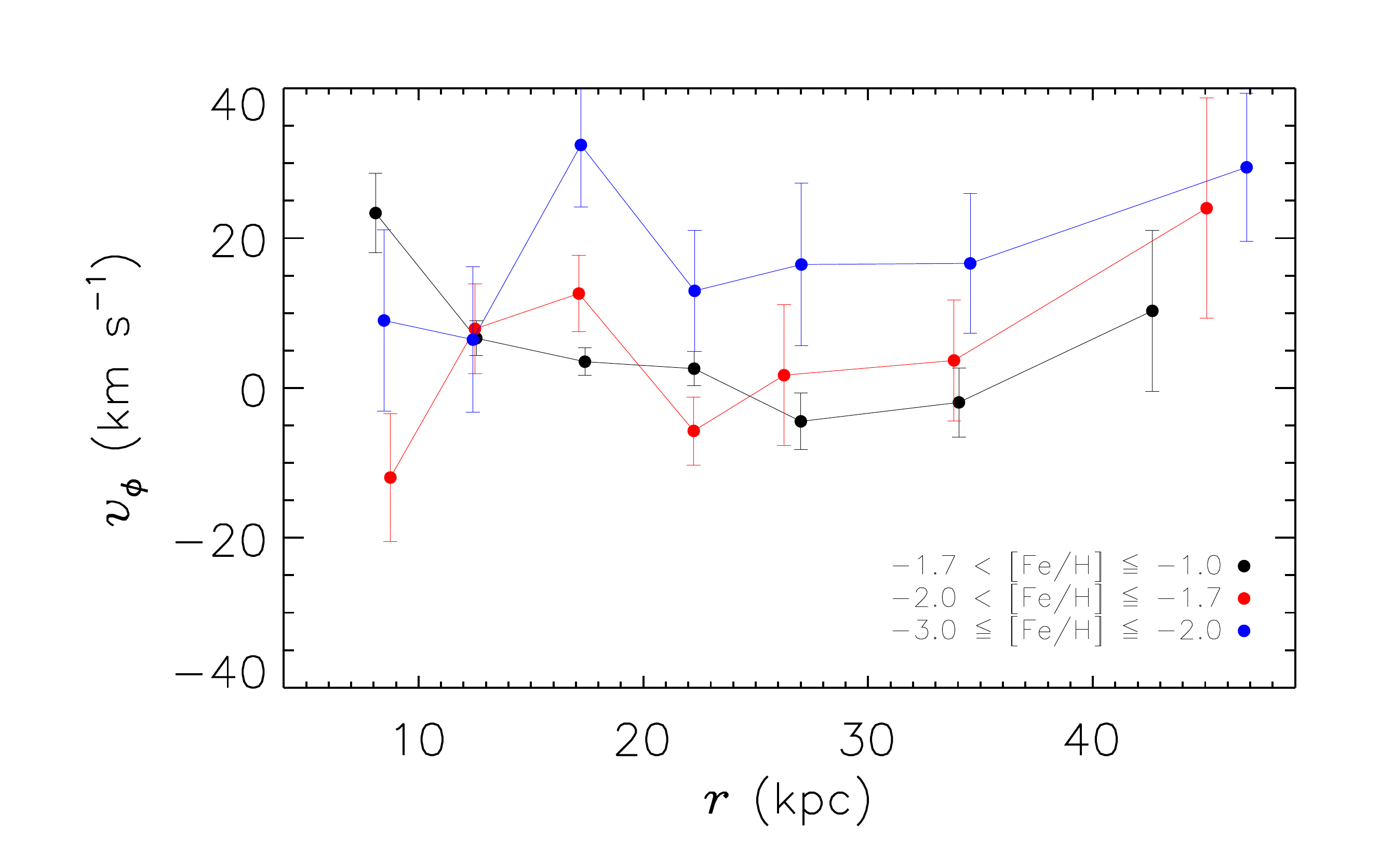}
    \caption{ Rotational velocity $v_{\phi}$ for three different metallicity populations ($- 1.7 \leq \rm{[Fe/H]} \leq -1.0$, $- 2.0 \leq \rm{[Fe/H]} \leq -1.7$, $- 3.0 \leq \rm{[Fe/H]} \leq -2.0$) as a function of Galactocentric distance $r$.}
   \label{vphi}
\end{figure*}

 \begin{figure*}
  \centering
  \includegraphics[scale=0.38, angle=0]{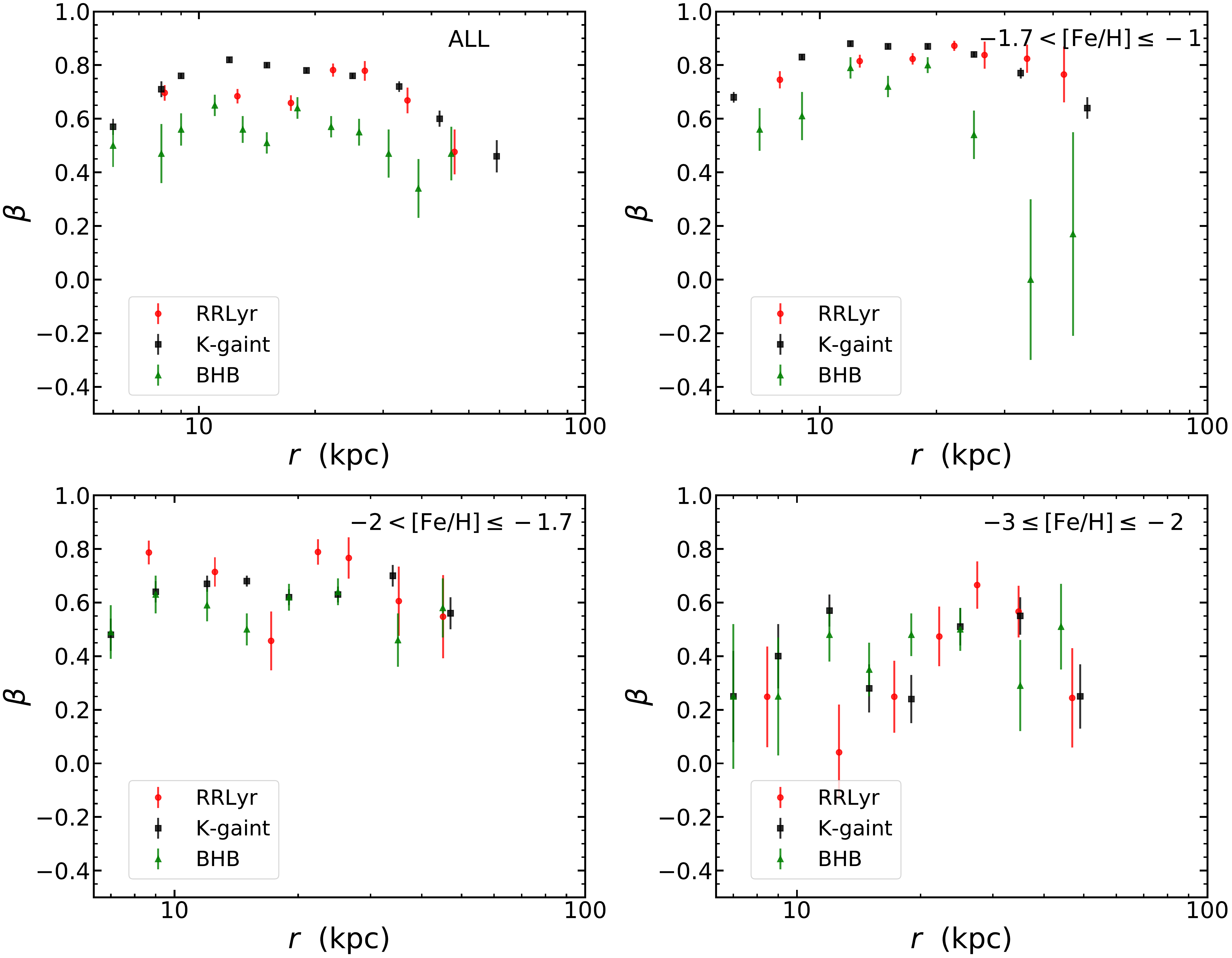}
    \caption{ Comparison of $\beta$ measured from our  RRL sample and from the K giants and BHB stars as presented by Bird et al. (2020). These samples represent the smooth, diffuse halo from which substructure has been removed by integrals of motion using the method of \citet[][in preparation]{Xue2022}. The top-left panel shows the anisotropy profile $\beta$ for all metallicities combined. The remaining three panels show the $\beta$ profile divided into different metallicity bins. }
  \label{figbetacomp}
\end{figure*}  
  
  \begin{figure*}
  \centering
  \includegraphics[scale=0.45, angle=0]{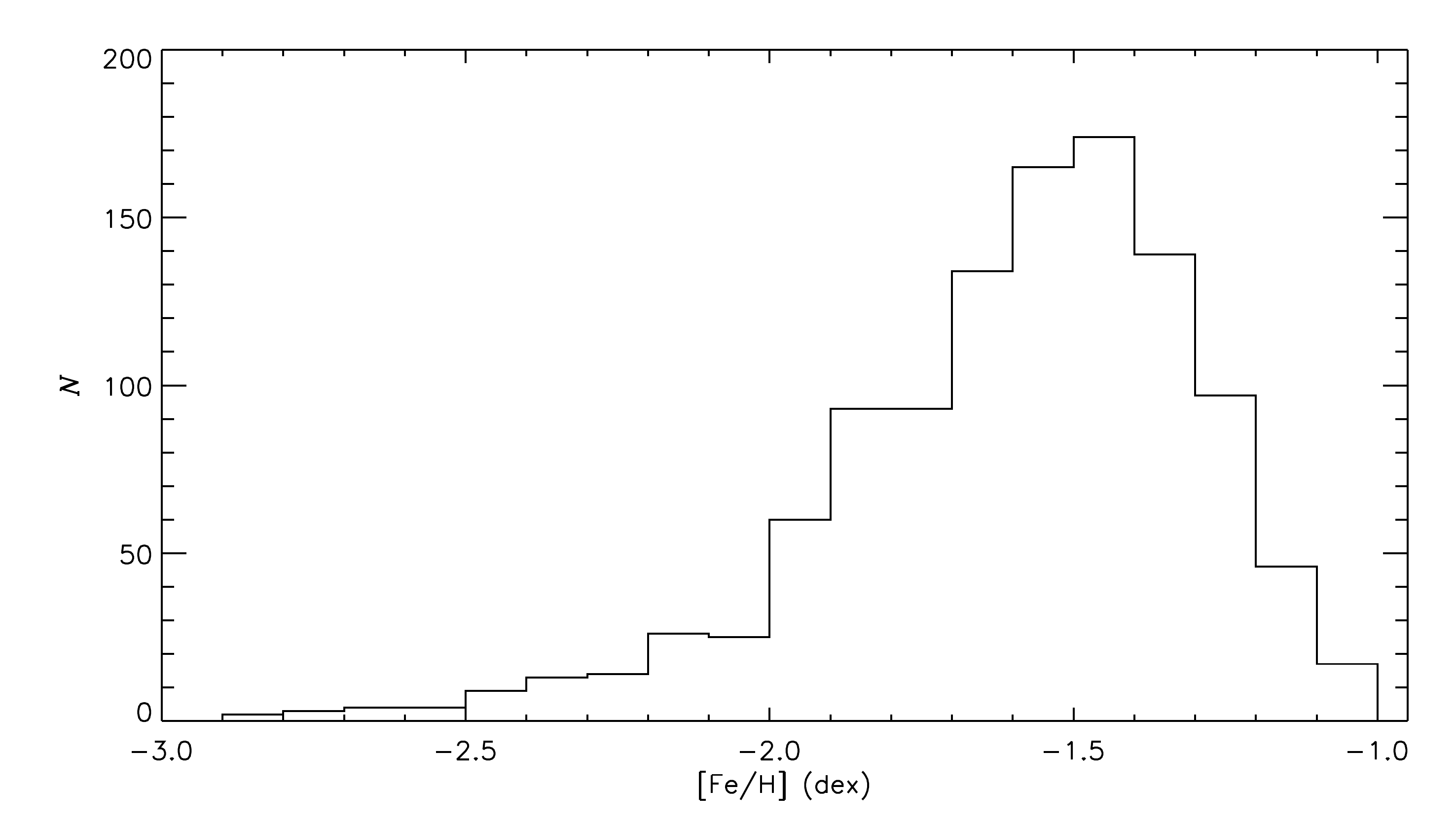}
    \caption{Metallicity distribution of the members of GES selected by integrals of motion using the method of \citet[][in preparation]{Xue2022}.}
   \label{gesmdf}
\end{figure*}  

 \begin{figure*}
  \centering
  \includegraphics[scale=0.6, angle=0]{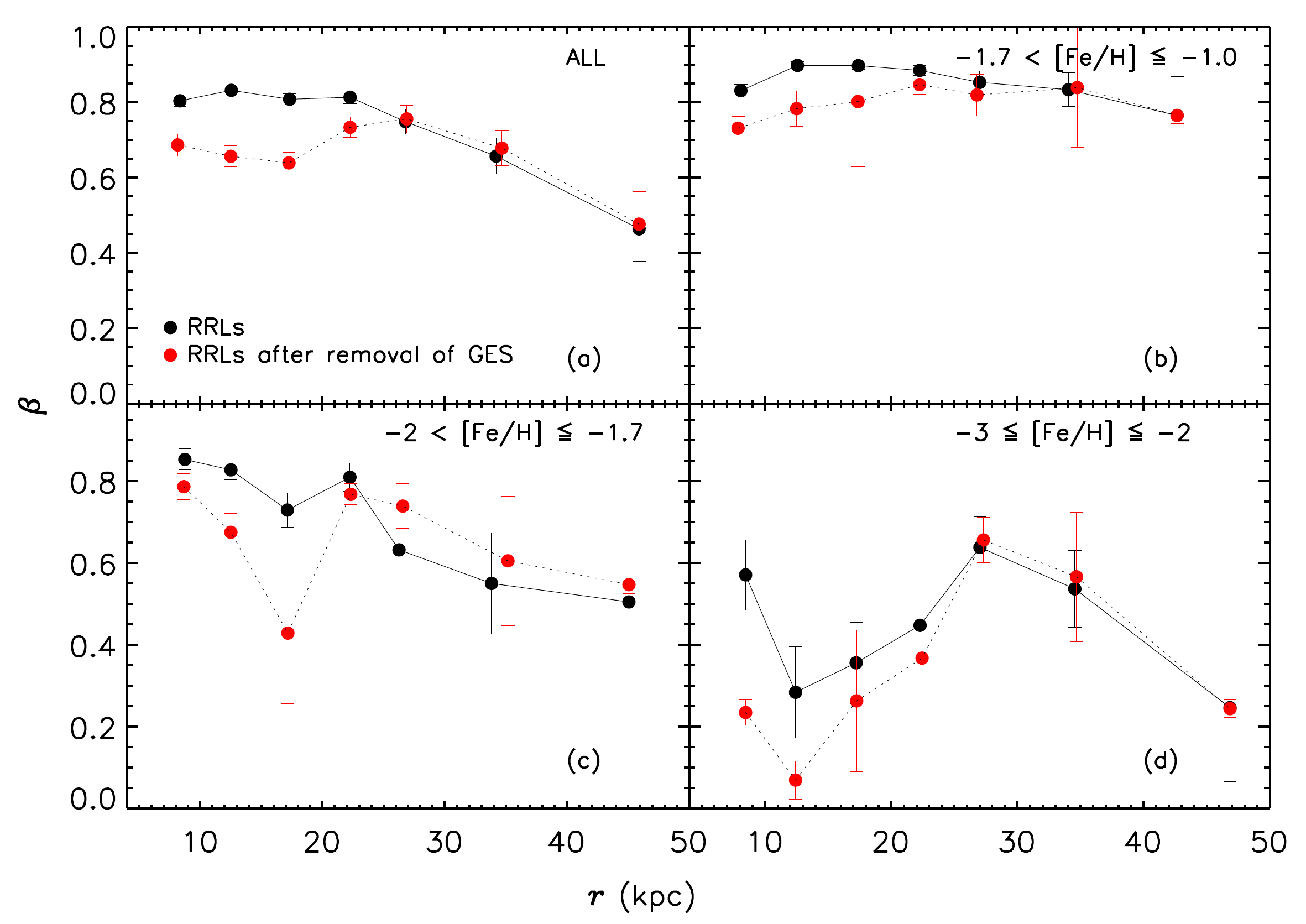}
    \caption{Comparison of  parameter $\beta$ for our RRL samples before (black) and after (red) the removal of GES as selected by integrals of motion using the method of \citet[][in preparation]{Xue2022}. Panel (a) shows the comparison for our total RRL sample and Panel (b), (c) and (d) show the results for our three metallicity populations. }
   \label{figbetayn}
\end{figure*}

The resulting mean metallicities of GES as a function of $r$ are shown in Figure\,\ref{fig_gradient}. 
A weak gradient is seen for this trend, although the uncertainties is considerable. To quantitively describe this trend, a linear fit is applied, using the IDL MPFIT package \citep{Markwardt2009}.
The best fit value of the gradient is $-$0.009 dex\,kpc$^{-1}$ with a relatively large uncertainty of 0.011 dex\,kpc$^{-1}$. This is the first report to find that GES exhibits a negative metallicity gradient. Note that the outer two radial bins are not considered for the fit. The metallicity of  the remaining inner halo component shows a very weak negative gradient but shallower than that of GES by a 3 factor, which is consistent with our expectation of an isotropic halo.
By assuming the more metal-rich component of our double-Gaussian model fitting for the inner halo represents the relics of GES, we find that the fraction of GES stars in the inner halo are 51, 68, 65, 67 and 53 per cent, respectively for each radial $r$ bin. This is an upper limit since some metal-rich halo stars may have, for example, formed in situ or been kicked up from the disk \citep{Yan_Yepeng2020}. These estimates are roughly consistent with the independent results from \citet{Wuwenbo2022}.

It should be noted that within the inner-most region of the inner halo, there could be residual contamination from the thick disk, and in the outer-most bins of the inner halo there could be contamination from the minor mergers, clearly appearing in the bin of $30 < r \leq 40$ kpc. To account for the above effects, we test an alternative model by assuming a three-Gauassian component model and re-fit the MDFs in the inner halo. We detect a slightly more shallow negative gradient of $-0.006 \pm 0.009$\,dex\,kpc$^{-1}$. From our two models tested for the inner halo MDFs, we detect a weak but non-zero negative metallicity gradient for GES as revealed by our RRL sample.
Meanwhile, we note the nil gradient can not be fully ruled out, given the large measurement uncertainties. 

 Many studies show that the Galactic stellar halo is largely unmixed and the inner halo is swamped with metal-rich tidal debris on highly radial orbits from an ancient, head-on collision, known as the GES \citep{Belokurov2018, Helmi2018}, and these mostly metal-rich stars are mixed with a more metal-poor and isotropic halo component built up from a superposition of various minor mergers \citep{Myeong2018.475}. These studies find that the metal-poor population is characterized by velocity isotropy (isotropic halo component) and the metal-rich population is characteristic of the relic stars of the GES.  \citet{Iorio2019, Iorio2021} also demonstrate that the halo RRL can be described as a superposition of an isotropic and radially-biased halo component.  They find that the radially-biased portion of the halo is characterized by a high orbital anisotropy ($\beta \sim $0.9)  and contributes between 50 and 80 per cent of the halo RRL at $5 < r < 25$ kpc. \citet{Lancaster2019} demonstrate that GES accounts for at least 50 per cent at  $r < 25$ kpc by using BHB stars. Comparing with $N$-body simulations, \citet{Naidu2021} also find that the GES merger event delivered about half of the Galactic stellar halo. \citet{Wuwenbo2022}  explore the contribution of GES to the Galactic halo by applying a Gaussian mixture model to K giant and BHB samples and find that GES contributes about 41 $-$ 74 per cent for the inner stellar halo ( $r < $30 kpc). All these agree well with our finding based solely on the metallicity density map and MDFs.
  
  In the following section, we investigate whether we see such results in kinematic space as seen in chemical space using our RRL sample.

\subsection {Velocity dispersion}

Motivated by the work of \citet{Bird2020}, we divide our RRL sample stars into three populations according to their metallicities, namely, a metal-rich population ($- 1.7 < \rm{[Fe/H]} \leq -1.0$), a metal-medium population ($- 2.0 < \rm{[Fe/H]} \leq -1.7$) and a metal-poor population ($- 3.0 < \rm{[Fe/H]} \leq -2.0$). We adopt the bootstrap resampling technique to calculate the mean velocity dispersion and their errors in spherical coordinates (i.e., radial $\sigma_r$, azimuthal $\sigma_ {\theta}$ and rotational $\sigma_\phi$) for the same Galactocentric radial bins as used in Section\,\ref{subsec:metallicity} and Figure\,\ref{fig5}. We estimate the velocity dispersion independently for each velocity component. To do so, we randomly select 80 per cent of the stars for each bin to calculate their mean velocity and corresponding dispersion and repeat the process 1,000 times. The estimate of the velocity dispersion $\sigma_{\rm obs}$ and its error in each radial bin is taken from the mean and standard deviation of the yielded distribution by the bootstrap technique. 
Finally, the intrinsic velocity dispersion $\sigma$ in each radial bin is obtained by subtracting the mean velocity uncertainty $<v_{\rm err}>$ of the concerned radial bin from the observed velocity dispersion estimated by the bootstrap technique, i.e. $\sigma = \sqrt{\sigma_{\rm obs}^2 - <v_{\rm err}>^2}$.

Figure\,\ref{fig6} shows the variation of velocity dispersion with $r$ for these three metallicity populations.  We find that the values of $\sigma_{r}$ are extensively larger than the corresponding $\sigma_\theta$ and $\sigma_\phi$ for the metal-rich and metal-medium populations, but the values of $\sigma_r$ are comparable with those of $\sigma_\theta$ and $\sigma_\phi$ for the metal-poor population at $r < 30$ kpc. At $r \geq 30$ kpc, the three components of the velocity dispersion ($\sigma_r$, $\sigma_ \theta$, $\sigma_ \phi$) are all comparable ($\sigma_r$ remaining slightly dominant), which is consist with that found by \citet{Bird2019, Bird2020}. We speculate that the strong radial profile for metal-rich and metal-medium populations of the inner halo may be caused by  GES as described by \citet{Belokurov2018}.

\subsection {Anisotropy parameter}
\label{subsection:beta}

  The velocity anisotropy parameter $\beta$, defined as $\beta = 1- (\sigma_\theta^{2}+\sigma_\phi^{2})/(2\sigma_r^{2})$, aids in characterizing the shape of the velocity ellipsoid as predominantly radial or tangential, or isotropic. Stellar systems with predominantly radial orbits yield $\beta > 0$. Tangential orbits yield $\beta < 0$. Isotropy orbits yield $\beta = 0$. By definition, $\beta \leq 1$. With the release of {\it Gaia} DR2, a large number of high quality stellar proper motions became available. With these in combination with radial velocity and distance measurements, $\beta$ can be measured directly over large Galactocentric distance. The first extensive, directly measured $\beta$ profile was published by \citet{Bird2019}, who profile the velocity dispersion in the stellar halo using a sample of 8600 K-giant stars from LAMOST DR5 \citep{Cui2012}. Incorporating the more accurate distances of BHB stars, \citet{Bird2020} redetermined the velocity anisotropy to 100 kpc in Galactocentric distance. 
   
   RRL stars, as an independent probe to explore the Galactic halo, also are suitable to measure the velocity anisotropy parameter $\beta$. We obtain proper motions of these stars from {\it Gaia} EDR3 \citep{GaiaCollaboration2021}. The values of $\beta$ and its errors are calculated from the velocity dispersions (and uncertainties) derived in the above section. Figure\,\ref{fig7} shows the anisotropy parameter $\beta$ as a function of Galactocentric radius for the three metallicity populations. Obviously, the metallicity and anisotropy are correlated. For the inner halo, we find that the metal-rich population exhibits extreme radial anisotropy ($\beta \sim 0.9$) and the value is nearly constant with Galactocentric radius, the metal-medium population also exhibits strong radial anisotropy ($0.6 \leq \beta \leq 0.8$), which reflects some contributions of GES. The metal-poor population shows less radial anisotropy ( $0.2 \leq \beta \leq 0.6$) which indicates that this more isotropic halo component is built up from a superposition of various minor mergers. The values are $\beta \sim 0.6$ at 25 \,kpc $\leq r \leq$ 30 \,kpc for both metal-medium and metal-poor populations. In the outer halo at $r >$ 30 kpc, the metal-medium and metal-poor populations exhibit similar mildly radial anisotropy ($\beta \sim$ 0.5). In agreement with the velocity dispersion profiles of Figure\,\ref{fig6}, the trends in the $\beta$ profiles exhibit change at $r \approx 30$ kpc. This also indicates $r \sim 30$ kpc may be the break radius between the inner and outer halo, which is highly consistent with the results from our chemical analysis. For the metal-medium  populations, we see a significant dip at $r \sim$ 17 kpc. Previous work noted and provided various explanations on the origin of the ``dip". For example, \citet{Kafle2012} attribute the origins of the ``dip" to an undetected feature in the Galactic potential or to a transition between two parts of the Galactic halo. \citet{King2015} also find a tangential ``dip" and suggest that it might be caused by substructures (Sagittarius stream or other streams within their sample). \citet{Bird2015} suggest the ``dip" is transient. \citet{Loebman2018} analyze $\beta$ from simulations and conclude that dips in $\beta$ might indicate a satellite passing or infalling.  Excitingly, as shown by Figure\,\ref{subs}, a significant peak is detected in $v_{\theta} = 50$\,km\,s$^{-1}$ in this radial bin. We speculate that a new substructure is located around 17 kpc with $\rm{[Fe/H]} \leq -1.7$ based on our sample, but this needs to be checked further and we leave this as future work. It is worth mentioning that there is a ``wide dip” for the metal-poor population in the radial bin $10 < r < 20$\,kpc, but a clear conclusion is hampered by our limited sample size and large uncertainties on the beta profile.

  \subsection {Velocity distribution ($v_{r}$, $v_{\phi}$)}
  We also show the radial and rotational velocity $v_{r}$ - $v_{\phi}$ distribution for the inner and outer halo and its dependence on metallicity in Figure\,\ref{fig:fig7}. For the inner halo ($r \leq 30$ kpc), the large radial velocity dispersion of the metal-rich population makes the $v_{r}$ - $v_{\phi}$ distribution appear strongly as a ``sausage." The sausage-like morphology becomes weaker and more rounded with decreasing metallicity. The $v_{r}$ - $v_{\phi}$ distribution for the outer halo ($r \geq 30$ kpc) is similar with that of the metal-poor population of the inner halo with a round, isotropic shape. The highly radial, metal rich component in the inner halo is direct evidence of the ancient satellite GES, that received its namesake from its sausage-like $v_{r}$ -$v_{\phi}$ distribution \citep{Belokurov2018}. 
  
  \citet{Lancaster2019} and \citet{Necib2019} found that the metal-rich part of the halo has a double-peaked distribution in $v_{r}$. \citet{Iorio2021} further showed that the distance between the two peaks is a function of $r$ and the double-peaked distribution disappears at r $>$ 15 $\sim$ 20 kpc. As checked the distribution of $v_r$ of our sample, we find similar conclusions as found by \citet{Iorio2021}.

There is still hot debate about the presence of net rotation for the halo stars. For example, \citet{Tian2019} report a slightly prograde rotation of the local halo. \citet{Utkin2018} find a weak prograde rotation in the inner halo and a retrograde rotation in the outer halo. \citet{Iorio2021} find a weak retrograde rotation for both the isotropic and radially biased components.  Figure \ref{vphi} shows the median azimuthal velocity as a function of $r$ for the three metallicity populations of our RRL sample. In general, no apparent rotation is found for the metal-rich and metal-medium populations, but a weak prograde rotation ($\sim$ 16.5 \kms) is clearly seen for the metal-poor population.
 
\section{Discussion}
  \label{discussion}
   \subsection{Comparison with \citet{Bird2020}}
  \citet{Bird2020} use both K giants and BHB stars to analyze the anisotropy profile of the stellar halo of the Milky Way. They remove substructure from their sample by using the method developed by \citet[][in preparation]{Xue2022}. To facilitate the discussion in this section, we use the same method to remove the substructures, namely, we utilize the friends-of-friends algorithm with the separation between two stars defined in integrals-of-motion space to identify substructures (for more details, please refer to \citetalias{Wang2022}). We find that the large majority of stars (1,118 out of the total 1,502 stars identified as substructure) are members of GES.

Figure\,\ref{figbetacomp} shows the comparisons of the smooth diffuse halo samples after removal of substructure. The top-left panel shows the anisotropy profile $\beta$ measured from our RRL sample and from the K giants and BHB stars for all metallicities combined. For $r \leq 20$ kpc, the $\beta$ profile of RRL is located between those of the K giants and BHB stars. However,  the $\beta$ value for RRL is very similar with that of the K giants when the Galactocentric distance is $>$ 20 kpc. The other three panels show the $\beta$ profiles for metal-rich, metal-medium, and metal-poor populations. We can find that the results from our RRL stars are consistent with those of the K giants for the metal-rich population. For the metal-medium and metal-poor populations, RRL, K giants and BHBs are basically consistent with each other within 3-$\sigma$ uncertainty.
The same conclusion with \citet{Bird2020} can be drawn that the anisotropy profile  depends on metallicity. As shown by figure 1 of \citet{Bird2020}, the mean metallicity of the K giants is richer than that of the BHB stars. The peak value [Fe/H]$ = -1.5$ of the metallicity distribution for our RRL sample is very close to that of the K giants ( [Fe/H] $\sim -$1.4), thus the similarity between the metal-rich anisotropy profiles for RRL and K giants is reasonable.   
  
    \subsection{ Properties of Galactic halo after removal of GES} 
  GES, as the last major merger experienced by the Milky Way about 10 Gyr ago, seems to be responsible for a large fraction of halo stars in the inner halo. In Section
 \ref{subsec:metallicity}, we use two Gaussian components to fit the MDF for the inner halo (five radial bins) and find that the upper limit of GES contributions for $r < 30$ kpc are 51, 68, 65, 67 and 53 per cent. \citet[][in preparation]{Xue2022} identify substructure in integral-of-motion space, such that stars with similar integrals of motion also have similar orbits and the orbit of a star can be characterized by energy $E$ and the angular momentum vector $\vec{L}$. This method finds multiple groups defined as substructure belonging to the inner halo, many of which likely belong to the same large merged satellite GES. As described by \citetalias{Wang2022}, we utilize the following criteria to select the member stars of GES from the total sample of substructure member stars found by the \citet[][in preparation]{Xue2022}: (i) $-800<L_z<620\,\rm km\,s^{-1}\,kpc$, (ii) $-1.4\times10^5<E<-0.5\times10^5\,\rm km^2\,s^{-2}$, (iii) $L_{\perp}=\sqrt{L_x^2+L_y^2}<3500\,\rm km\,s^{-1}\,kpc$. In total we obtain 1,118 member stars of GES. The metallicity distribution of those members is shown in Figure\,\ref{gesmdf}, which is similar to the distribution of GES shown in Figure\,\ref{fig5}. This method imposes rigorous restrictions on the candidates in terms of energy and angular momentum. The number of selected stars by this method for a substructure are largely less than the total number of member stars in this substructure. The GES contributions selected by this method for our five defined radial bins in the inner halo are  20 per cent (125/619), 29 per cent (313/1085), 32 per cent (337/1059), 29 per cent (212/722) and 28 per cent (100/360), which may represent the lower limits of GES contributions. The contribution of GES members for $r > 30$ kpc is only 6 per cent (31/520).
 
  We compare the anisotropy profile for samples before and after the removal of GES in Figure \ref{figbetayn}. For the whole sample, we find that GES mainly influences the inner halo ($r \leq 30$ kpc) and has almost no effect on the outer halo. After the removal of GES, the anisotropy parameter $\beta$ drops by $\sim 0.1$ for the metal-rich population of the inner halo, but shows almost no change for the outer halo.  For the metal-medium population, the anisotropy $\beta$ profile drops down by $\sim 0.1$ for inner halo and the significant dip near $r\sim17$ (discussed in Section \ref{subsection:beta}) remains. The anisotropy profile of the metal-medium population slightly increases in outer halo with quite large uncertainties. 
  For the metal-poor population, the anisotropy parameter $\beta$ slightly decreases by $\sim 0.1$
  in the inner halo but shows almost no change in the outer halo.
  This indicates that GES is mainly composed of metal-rich stars with extremely radially dominated orbits and is distributed in the inner halo but still contains some metal-medium and metal-poor stars on less radial orbits. By comparing with the metal-poor sample of Figure \ref{fig7} (blue curve), 
  We find the metal-poor stars after removing GES remain on radially dominated orbits  with some oscillation in the anisotropy profile. We speculate that this profile may be caused by \textbf{hierarchical assembly involving the plethora of discovered substructures and merger remnants} such as the Helmi streams \citep{Helmi1999Nature}, Virgo overdensity \citep{Vivas2004}, GD-1 \citep{Grillmair2006, Grillmair2016}, Hercules Aquila Cloud \citep{Belokurov2007}, Sequoia \citep{Myeong2019}, and Thamnos \citep{Matsuno2019}.
       
 \section{Conclusion}
Using 4,365 RRL with 7D information compiled by \citetalias{Liu2020} together with an extended data set obtained by the same method, we present in detail the chemical and kinematic properties of the Galactic halo. Both the chemical and kinematic results suggest that the Galactic halo has two distinct components, the inner halo and outer halo, with a break radius around 30 kpc, similar to the previous break radius measured from the density distribution of halo stars. The inner halo is dominated by a metal-rich component $ -1.8 < \rm[Fe/H] <-1.1$ located at $r$ between 7 and 27 kpc with extremely radially eccentric orbits and $\beta$ reaching values as high as 0.9.  We find that this component dominates $\sim$ 60 per cent of the inner halo. Its MDF, velocity anisotropy profile, and velocity distribution are all consistent with the characteristics of the GES described by many recent works. We discovered for the first time that GES shows a slightly negative metallicity gradient. The metal-poor population of the inner halo is characterized as a long tail in the MDF and by a value of  anisotropy $\beta$ around 0.5, which is similar to that of the outer halo. The MDF of the outer halo is very broad with several weak peaks and the value of $\beta$ is around 0.5 for all metallicities. We find that after the removal of GES, the main differences characterizing the inner and outer halos (MDF, velocity anisotropy profile, and velocity distribution) remain evident. Further investigation is needed to determine the processes which induce the distinct components of the Galactic stellar halo.

\section*{Acknowledgements}

We thank the anonymous referee for his$/$her constructive comments. This work is supported by the Key Laboratory Fund of Ministry of Education under grant No. QLPL2022P01, the National Science Foundation of People's Republic of China (NSFC) with No. U1731108 and National Key R$\&$ D Program of China No. 2019YFA0405500. YH acknowledges the National Key R$\&$D Program of China No. 2019YFA0405503 and the NSFC with Nos. 11903027 and 11833006. HWZ and FW acknowledge the National Key R$\&$D Program of China No. 2019YFA0405504 and the NSFC with Nos. 2019YFA0405504, 11973001, 12090040 and 12090044. HJT acknowledges NSFC with No. 11873034 and the Hubei Provincial Outstanding Youth Fund No. 2019CFA087. We acknowledge the science research grants from the People's Republic of China Manned Space Project under grants CMS-CSST-2021-B05, CMS-CSST-2021-A08, CMS-CSST-2021-B03, and China Space Station Telescope Milky Way and Nearby Galaxies Survey on Dust and Extinction Project CMS-CSST-2021-A09. The Guoshoujing Telescope (the Large Sky Area Multi-Object Fiber Spectroscopic Telescope LAMOST) is a National Major Scientific Project built by the Chinese Academy of Sciences. Funding for the project has been provided by the National Development and Reform Commission. LAMOST is operated and managed by the National Astronomical Observatories, Chinese Academy of Sciences. This work has made use of data from the European Space Agency (ESA) mission {\it Gaia} (\url{https://www.cosmos.esa.int/gaia}), processed by the {\it Gaia} Data Processing and Analysis Consortium (DPAC, \url{https://www.cosmos.esa.int/web/gaia/dpac/consortium}). Funding for the DPAC has been provided by national institutions, in particular the institutions participating in the {\it Gaia} Multilateral Agreement. This work also has made use of data products from the SDSS, 2MASS, and WISE.

%\appendix
%\section{Priors of the parameters of the metallicity distribution model}

%The priors of the assumed parameters of the metallicity [Fe/H] distribution model of each radial bin are all uniform within the ranges listed in Table\,A1.

%%%%%%%%%%%%%%%%%%%%%%%%%%%%%%%%%%%%%%%%%%%%%%%%%%
\section*{Data Availability}

The data supporting this article will be shared upon reasonable request sent to the corresponding authors.



\bibliographystyle{mnras}
\bibliography{bibfile} % if your bibtex file is called example.bib

%\begin{thebibliography}{99}
%\end{thebibliography}

%%%%%%%%%%%%%%%%%%%%%%%%%%%%%%%%%%%%%%%%%%%%%%%%%%

%%%%%%%%%%%%%%%%% APPENDICES %%%%%%%%%%%%%%%%%%%%%

%%%%%%%%%%%%%%%%%%%%%%%%%%%%%%%%%%%%%%%%%%%%%%%%%%

% Don't change these lines
\bsp	% typesetting comment
\label{lastpage}
\end{document}